\documentclass[twocolumn,aps,pra,nofootinbib,groupedaddress,amsmath,amssymb,longbibliography]{revtex4-1}
\usepackage{graphicx}
\usepackage{bm}
\usepackage[colorlinks=true,	linkcolor=blue,urlcolor=blue,anchorcolor=blue,citecolor=blue,bookmarksnumbered]{hyperref}
\usepackage{float}
\usepackage[mathscr]{euscript}
\usepackage[english]{babel}
\usepackage{natbib}

\begin{document}
	\title{Superradiance induced multistability in one-dimensional driven Rydberg lattice gases}
	\author{Yunhui He$^{\star,1}$}
	\author{Zhengyang Bai$^{\star,2,4}$}
	\author{Yuechun Jiao$^{1,4}$}
	\author{Jianming Zhao$^{1,4}$}
	\email{zhaojm@sxu.edu.cn}
	\author{Weibin Li$^{3}$}
	\email{weibin.li@nottingham.ac.uk}
	\affiliation{$^1$State Key Laboratory of Quantum Optics and Quantum Optics Devices, Institute of Laser Spectroscopy, Shanxi University, Taiyuan 030006, China\\
		$^2$State Key Laboratory of Precision Spectroscopy, East China Normal University, Shanghai 200062, China\\
		$^3$School of Physics and Astronomy, and Centre for the Mathematics and Theoretical Physics of Quantum Non-equilibrium Systems, University of Nottingham, Nottingham, NG7 2RD, UK\\
		$^4$Collaborative Innovation Center of Extreme Optics, Shanxi University, Taiyuan 030006, China}
	
	\def\thefootnote{$\star$}\footnotetext{These two authors contributed equally to this work.}\def\thefootnote{\arabic{footnote}}
	
	\begin{abstract}
		We study steady state phases of a one-dimensional array of Rydberg atoms coupled by a microwave (MW) field where the higher energy Rydberg state decays to the lower energy one via single-body and collective (superradiant) decay. Using mean-field approaches, we examine the interplay among the MW coupling, intra-state van der Waals (vdW) interaction, and single-body and collective dissipation between Rydberg states. A linear stability analysis reveals that a series of phases, including uniform, antiferromagnetic, oscillatory, and bistable and multistable phases can be obtained. Without the vdW interaction, only uniform phases are found. In the presence of the vdW interaction, multistable solutions are enhanced when increasing the strength of the superradiant decay rate. Our numerical simulations show that the bistable and multistable phases are stabilized by superradiance in a long chain. The critical point between the uniform and multistable phases and its scaling with the atom number is obtained. Through numerically solving the master equation of a finite chain, we show that the mean-field multistable phase could be characterized by expectation values of Rydberg populations and two-body correlations between Rydberg atoms in different sites.
	\end{abstract}
	
	\maketitle

	\section{Introduction}
	Collective behaviors are intriguing in various many-body systems and attract intensive interest currently. Among them, superradiance is a cooperative radiation effect in dense atomic samples~\cite{gross1982Superradiance}. Spontaneous decay of individual atoms occurs due to fluctuations of vacuum fields surrounding atoms. When interatomic separation $R_{jk}$ is smaller than wavelength $\lambda$ of the respective transition, i.e. the Dicke limit $R_{jk} \ll \lambda$~\cite{dicke1954coherence}, decay becomes collective such that its rate depends on the number of atoms in the ensemble, and hence can be much larger than the individual decay rate~\cite{ficek2002entangled}. Since predicted by Dicke, superradiance has been confirmed in a variety experimental settings including Rydberg atoms~\cite{gross1979maser, Moi1983Rydberg, Wang2007superradiance, Grimes2017Direct, hao2021observation}, cavities~\cite{kaluzny1983observation, mlynek2014observation, suarez2022superradiance}, Bose-Einstein condensates~\cite{inouye1999superradiant, lode2017fragmented, chen2018experimental}, and quantum dots~\cite{scheibner2007superradiance}. On the other hand, insights gained from the study of superradiance allow us to develop applications in quantum metrology~\cite{wang2014heisenberg, liao2015gravitational}, narrow linewidth lasers~\cite{haake1993superradiant, bohnet2012a, norcia2016cold} and atomic clocks~\cite{norcia2016superradiance}, etc.
	
	Rydberg atoms become an ideal platform for studying superradiance because of their millimeter-wavelength energy intervals, inherent dissipation~\cite{Moi1983Rydberg, Gallagher1994Rydberg}, and spatial configurability~\cite{browaeys_many-body_2020-1,scholl_quantum_2021,ebadi_quantum_2021}. Rydberg atoms have extremely large electric dipole transition moments that can cause strong and long-range interactions of Rydberg states. There have been numerous theoretical and experimental investigations on the competition between dissipation and strong Rydberg atom interactions~\cite{Weimer2008quantum, lesanovsky_kinetic_2013,marcuzzi_universal_2014,Malossi2014Full,hoening_antiferromagnetic_2014,sibalic_driven-dissipative_2016,Letscher2017Bistability, Gutierrez2017Experimental,yan_electromagnetically_2020,ding_phase_2020-2}. The strong interaction between Rydberg atoms leads to blockade effects~\cite{Lukin2001Dipole, Tong2004Local, Singer2004Suppression, Heidemann2007Evidence}. Taking into account of single-body dissipation, novel phases~\cite{lee2011antiferromagnetic, lee2012collective, Hu2013Spatial} and critical behaviors~\cite{Weimer2008quantum, tomadin2011nonequilibrium, Zimmermann2018Wiseman, Hannukainen2018dissipation, Ferreira2019Lipkin} emerge in such driven-dissipation many-body setting. We have recently experimentally  observed blackbody radiation enhanced superradiance of ultracold Rydberg atoms in a magneto-optical trap (MOT)~\cite{hao2021observation}. In a cold gas of dense Rydberg atoms, decay from $|nD\rangle$ state to $|(n+1)P\rangle$ state is much faster than the single-body decay rate, which is identified to be superradiant. It is found that the strong van der Waals (vdW) interaction between Rydberg atoms plays crucial roles. The interplay between superradiance and vdW interactions affects the many-body dynamics as well as scaling of the superradiance with respect to number $N$ of Rydberg atoms.
	\begin{figure}
		\centering
		\includegraphics[width=1\linewidth]{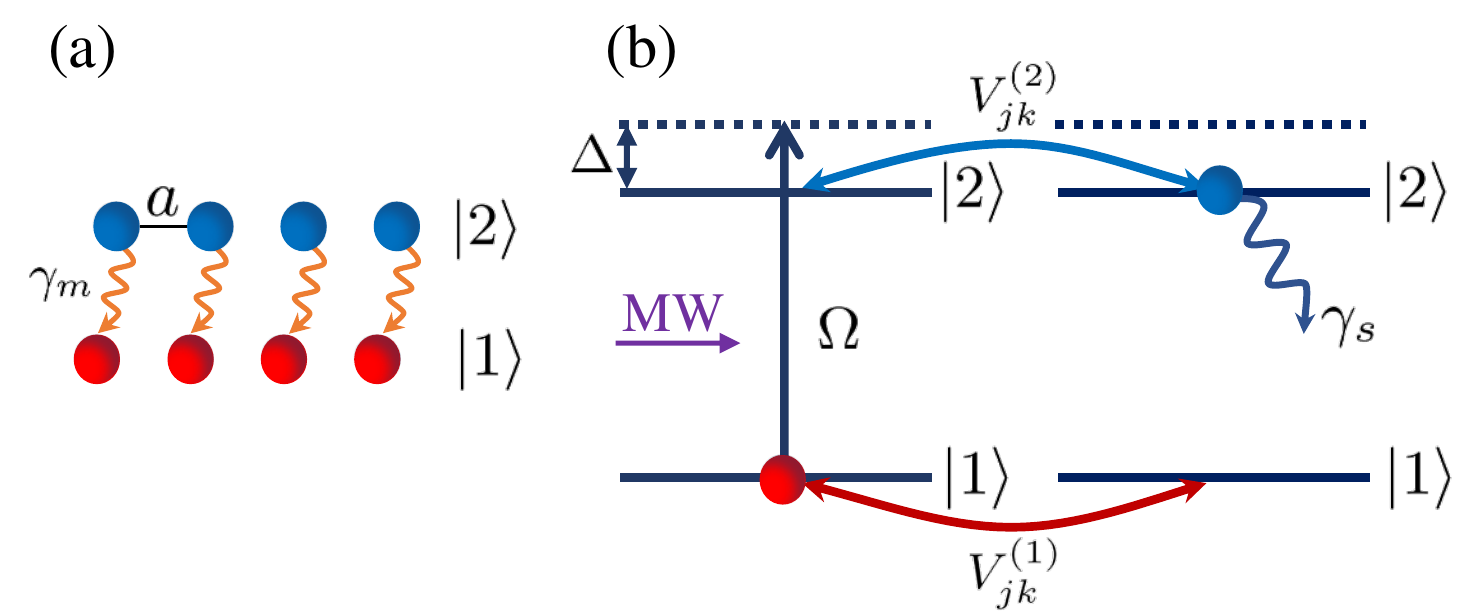}
		\caption{1D array of interacting Rydberg atoms. (a) A large number of atoms are fixed on the 1D lattice. $a$ is the lattice distance and $\gamma_m$ is the many-body decay term. (b) the energy level of our model. A microwave field with Rabi frequency $\Omega$ drives atoms from $|1\rangle$ to $|2\rangle$. $\Delta = \omega - \omega_0$ is the detuning from the two-level transition $\omega_0$. $\gamma_s$ is the single-body decay term and $V_{jk}^{(1)}(V_{jk}^{(2)})$ is the vdW interaction between the same state $|1\rangle(|2\rangle)$ when atoms in site $j$ and $k$.}
		\label{fig:model}
	\end{figure}
	
	In this work, we study superradiance between two Rydberg states in a 1D lattice (see Fig.~\ref{fig:model}), where atoms experience strong vdW interactions and are coupled by a microwave field. This lattice setting allows us to explore superradiance between Rydberg states in a controllable fashion, e.g. modifying the effective collective dissipation and interaction strength between Rydberg atoms by changing the atomic density and principal quantum number. Dynamics of the driven-dissipative Rydberg lattice is governed by a Lindblad master equation. We first establish mean-field phase diagrams as a function of external drive and detuning. We find a variety of stationary phases, including antiferromagnetic, oscillatory, phase bistabilities, and multistabilities. We show that Rydberg superradiance leads to multistable phases that are absent in previous studies~\cite{Lee2014dissipative}. In a finite chain, we obtain steady states by numerically solving the master equation.  Two-body correlations and Rydberg populations exhibit different features in the corresponding mean-field phases, and could signify the emergence of bistable and multistable phases.
	
	The paper is organized as follows. In Sec.~\ref{model and methods}, we describe master equation of the Rydberg atoms on a 1D lattice. In Sec.~\ref{Mean field phase diagrams}, we use mean-field theory and ansatz to analyze steady states of the model. Different phases, described by order parameter $S_z$, are identified. We show dependence of the steady state phase diagrams on the collective (nonlocal) dissipation. In Sec.~\ref{Analysis of mean field phases}, we explore the linear stability of the steady state. Dynamics of different phases, in particular the multistable phases, are discussed. In Sec.~\ref{Quantum many-body dynamics of finite 1D chains}, we numerically obtain the quantum correlation and Von Neumann entropy in the quantum master equation, and link the result with mean-field predictions. We conclude in Sec.~\ref{discussion and conclusions}.
	
	\section{the Model}\label{model and methods}
	We consider a one-dimensional lattice of $N$ atoms in electronically high-lying Rydberg states $|1\rangle$ and $|2\rangle$, as depicted in Fig.~\ref{fig:model}. Similar to the experiment~\cite{hao2021observation}, we assume that states $|2\rangle=|nD\rangle$ and $|1\rangle=|(n+1)P\rangle$ with $n$ to be the principal quantum number. These states are coupled by a microwave (MW) field with Rabi frequency $\Omega$ and detuning $\Delta$. In Rydberg state $|s\rangle$ $(s=1,2)$, atoms located at site $j$ and $k$ interact strongly with vdW interactions $V_{jk}^{(s)} = C_6^{s}/a^6{|j-k|}^6$ where $C_6^s$ and $a$ are the dispersion coefficient and lattice constant. The interstate interaction is neglected, as the two states are energetically separated~\cite{olmos_amplifying_2011}. Hamiltonian of the many-body system is given by ($\hbar=1$)~\cite{hao2021observation}
	\begin{align}\label{eq:Ham}
		\begin{split}
			H&=\sum_j^N\left[\frac{\Omega}{2}\sigma_x^j-\frac{\Delta}{2}\sigma_z^j\right] + \frac{1}{2}\sum_{\alpha=1,2}\sum_{k\neq j}^NV_{jk}^{(\alpha)}\sigma_{\alpha\alpha}^j\sigma_{\alpha\alpha}^k\\
			&+\frac{1}{2}\sum_{k\neq j}^NV_{jk}^{DD}\!\left(\sigma_x^j\sigma_x^k+\sigma_y^j\sigma_y^k\right).
		\end{split}
	\end{align}
	where $\sigma_{\xi}^j$ ($\xi=x,y,z$) are the Pauli matrices on site $j$, $\sigma_{(\pm)}^j = \frac{1}{2}(\sigma_x^j \pm i \sigma_y^j)$ is the raising (lowering) operator, and $\sigma_{ss}^j = [1 +(-1)^s \sigma_z^j]/2$ are projection operators to the Rydberg state. The dipole-dipole (DD) interaction is given by  $V_{jk}^{DD} = C_3(1-3\cos^2\theta_{jk})/a^3|j-k|^3$, where  $\theta_{jk}$ is the angle between their internuclear axis and quantization axis.
	
	The Rydberg states are subject to individual and collective (superradiant) decay~\cite{hao2021observation}.
	Dynamics of the system are governed by the Lindblad master equation~\cite{ficek2002entangled}
	\begin{equation}\label{eq:master}
		\dot{\rho}(t)=-i[H,\rho(t)]+\mathcal{L}[\rho(t)],
	\end{equation}
	where $\rho$ is the many-body density matrix, and operator $\mathcal{L}(\rho)$ describes the dissipation,
	\begin{equation}\label{eqn:dissipa}
		\mathcal{L}[\rho]=\sum_{j,k}^N\frac{\Gamma_{jk}}{2}[2\sigma_-^j\,\rho\,\sigma_+^k-\{\sigma_+^k\sigma_-^j,\rho\}],
	\end{equation}
	where $\Gamma_{jk}$ is the collective decay rate between site $j$ and $k$. When $j=k$, single-body decay rate $\gamma_s=\Gamma_{jj}=\omega_j^3\mu_j^2/3\pi\epsilon_0\hbar c^3$, where $\omega_j$ is the transition frequency and $\mu_j$ is the dipole moment~\cite{ficek2002entangled}. If the atom separation $R_{jk}$ is much larger than the photon wavelength $\lambda = 2\pi c/\omega$, the decay is dominated by the individual (local) ones. For densely packed atoms, superradiance leads to nonlocal dissipation that varies with the distance between atoms~\cite{ficek2002entangled,Olmos2014steady,parmee2018phases}. In our analysis, we neglect the distance dependence as the average spacing ($\sim\mu $m) between Rydberg atoms is much smaller than the MW wavelength ($\sim$ mm). In a mesoscopic setting (tens to hundreds of atoms), the collective decay becomes all-to-all with equal strength, i.e.  $\Gamma_{jk}=\gamma_m$~\cite{Lee2014dissipative}.
	
	In the following discussions, the DD interaction will be neglected for the following reason. First, in our recent experiment~\cite{hao2021observation} it has been shown that superradiance in dense Rydberg gases is strongly affected by the van der Waals interactions while effects due to the DD interaction is not significant. This is due to the fact that dipolar interactions are long-ranged ($\sim R^{-3}$), but the vdW interaction is short-ranged ($\sim R^{-6}$). The vdW interaction can be stronger than the DD interaction at short distances (see Appendix \ref{experimental parameter} for illustrations). Second, one can turn off the DD interaction by adopting the magic angle (i.e., $1-3\cos^2\theta_{jk}=0$) in the one-dimensional model (see Appendix \ref{experimental parameter} for details). The influence of the DD interaction on Rydberg superradiant dynamics will be discussed elsewhere.
	
	\section{Mean-field phases}\label{Mean field phase diagrams}
	The Hilbert space of the Hamiltonian grows as $2^N$, while the dimension of the density matrix is $2^{2N}$. The computational complexity prevents us from numerically solving the many-body problems when $N > 10$ in typical computers. Due to the dissipation, we could employ the mean-field (MF) theory to analyze the steady state and dynamics. In the MF approximation, the many-body density matrix $\rho$ is decoupled into individual ones through $\hat{\rho} \approx \Pi_i$ $\hat{\rho}_i$. This decoupling essentially ignores quantum entanglement between different sites~\cite{diehl2010dynamical}. We obtain MF equations of motion of the spin expectation values~\cite{hao2021observation}
	
	\begin{subequations}\label{eq:MF}
		\begin{align}
			\dot{S_x^j}&=-\frac{\gamma_s}{2}S_x^j+\Delta S_y^j+\sum_{k\neq j}\left(S_y^j S_V^{jk}+\mathcal{F}_x^{jk}\right),\label{eq:MFa}\\
			\dot{S_y^j}&=-\frac{\gamma_s}{2}S_y^j-\Delta S_x^j-\Omega S_z^j-\sum_{k\neq j}\left(S_x^j S_V^{jk}- \mathcal{F}_y^{jk}\right),\label{eq:MFb}\\
			\dot{S_z^j}&=-\frac{\gamma_s}{2}(1+2S_z^j)+\Omega S_y^j-\gamma_m\sum_{k\neq j}\mathcal{D}^{jk},\label{eq:MFc}
		\end{align}
	\end{subequations}
	where $S_\xi^j = \frac{1}{2}\text{Tr}(\sigma_\xi^j\hat{\rho})$ are the expectation values of operator $\sigma^j_\xi$,  $\mathcal{D}^{jk}=S_x^j S_x^k+S_y^j S_y^k$ and $\mathcal{F}^{jk}_{\xi} = \gamma_mS_z^jS_{\xi}^k$. We have defined the site-dependent interaction term $S_V^{jk} =[ V_{jk}^{(1)}(1-2S_z^k)-V_{jk}^{(2)}(1+2S_z^k)]/4=[( V_{jk}^{(1)}-V_{jk}^{(2)}) -2S_z^k(V_{jk}^{(1)}+V_{jk}^{(2)})]/4$, which is dependent on the interaction strength and $S_z$. It shows that the nonlinear interaction will decrease when $V_{jk}^{(1)} \sim -V_{jk}^{(2)}$. Note that the vdW interaction decreases rapidly with spin separations ($\propto 1/a^6|j-k|^6$). In the coherent regime, the classical groundstate forms crystalline structures in the thermodynamic limit~\cite{von_boehm_devils_1979,lan_emergent_2015,schaus_crystallization_2015,lan_devils_2018}. The Rabi coupling, on the other hand, could melt the crystalline phase~\cite{weimer_two-stage_2010}. The vdW type interaction between Rydberg atoms means that the nearest-neighbor (NN) interaction is $p^6$ times of other long-range interactions (with atom separation $pa$ with $p\ge 2$). Typically the long-range tail of the vdW interaction leads to subtle details in the crystal melting~\cite{sela_dislocation-mediated_2011-1,petrosyan_two-dimensional_2013-1,lan_quantum_2016}. Following Ref.~\cite{Lee2014dissipative}, we will take into account of the NN interaction $V_{1(2)}=C_{1(2)}/a^6$ in the following analysis. Without losing generality, we will scale energy with respect to $\gamma_s$ in the numerical simulations, except in Sec. IV-C.
	
	At the mean-field level, the nonlocal, collective decay leads to nonlinear dissipative terms in the mean-field equations, while local decay leads to linear dissipative terms (see Eq.~(\ref{eq:MF})). The collective decay is all-to-all and independent of distance. Depending on the parameters, we find Rydberg populations in the MF steady state can have different distributions along the lattice. To characterize the phases, we will use $S_z$ as an order parameter, and identify uniform (UNI), and non-uniform solutions.
	
	
	\subsection{Uniform phases}
	The uniform phase corresponds to spatially homogeneous excitation of both Rydberg states. To obtain the uniform solution, one can find the fixed point through
	\begin{subequations}\label{eq:UNI}
		\begin{align}
			\dot{S_x}&=-\frac{\gamma_s}{2}S_x+\tilde{\Delta}S_y+\kappa S_zS_x,\label{eq:UNIa}\\
			\dot{S_y}&=-\frac{\gamma_s}{2}S_y-\tilde{\Delta}S_x-\Omega S_z+\kappa S_zS_y,\label{eq:UNIb}\\
			\dot{S_z}&=-\frac{\gamma_s}{2}(1+2S_z)+\Omega S_y-\kappa\left(S_x^2+S_y^2\right),\label{eq:UNIc}
		\end{align}
	\end{subequations}
	where $\tilde{\Delta}=\Delta+S_V$, and $\kappa=\left(N-1\right)\gamma_m$. Order parameter $S_z$ in the UNI phase satisfies
	\begin{equation}\label{eq:UNI-SZ}
		\tilde{\Delta}^2+\left(\frac{\gamma_s}{2}-\kappa S_z\right)^2+\frac{\Omega^2S_z}{2S_z+1}=0.   
	\end{equation}
	This is a nonlinear function of $S_z$, where analytical solutions are typically difficult to derive. In a special case, $V_1=-V_2=V$, an analytical solution can be obtained. The expression of the solution is lengthy and is given in Appendix~\ref{Uniform Analytical Solutions}. In general conditions, solutions in the uniform phase are obtained numerically. According to values of Rydberg excitation, we further divide the UNI phase into low-excitation phase (ULE phase) when $-1/2 < S_z < -1/4$ (i.e., the population on level $|2\rangle$ $S_{22}=0.5+Sz$; it satisfies $0<S_{22}<1/4$ ), and high-excitation phase (UHE phase) if $-1/4 < S_z < 1/2$ (i.e., $1/4< S_{22} <1$ ).
	\subsection{Non-uniform phases}
	Due to the NN interaction, we employ a bipartite sublattice ansatz to analyze the stationary states. Here two NN sites, labelled with $A$ and $B$, repeat their pattern periodically throughout the lattice. With this periodicity in mind, Eq.~(\ref{eq:MF}) is simplified to the following coupled equations of the $A-B$ sublattice,
		\begin{subequations}\label{eq:AB}
			\begin{align}
				\dot{S_x^A}&=-\frac{\gamma_s}{2}S_x^A+(\Delta+S_V^B)S_y^A+\frac{N\gamma_m\mathcal{C}_x}{2} S_z^A,\label{eq:ABa}\\
				\dot{S_y^A}&=-(\Delta+S_V^B)S_x^A-\frac{\gamma_s}{2}S_y^A+\frac{N\gamma_m\mathcal{C}_y-2\Omega}{2}  S_z^A,\label{eq:ABb}\\
				\dot{S_z^A}&=\Omega S_y^A-\frac{\gamma_s}{2}\!(1+\!2S_z^A)-\frac{N-2}{2}\gamma_m\!(S^{A}_{\perp})^2-\!\frac{N\!\gamma_m}{2}\mathcal{D}^{AB}\!,\label{eq:ABc}
			\end{align}
		\end{subequations}
	with $S^{\alpha}_{\perp} = \sqrt{{S_x^{\alpha}}^2\!+\!{S_y^{\alpha}}^2}$ ($\alpha = A, B$) is the projection of $\alpha$-spin on the $x-y$ plane and $\mathcal{C}_{\xi} = (N-2)S_{\xi}^A/N+S_{\xi}^B$. Equations for $B$ sites can be obtained by swapping index $A$ and $B$ in Eq.~(\ref{eq:AB}). We then obtain MF steady state solutions by solving these equations numerically.
	
	According to values of $S_z^{\alpha}$, we identify antiferromagnetic (AFM), oscillatory (OSC) phase, and bistable/multistable phases. In AFM phases one sublattice has a higher excitation than the other ($S_z^A\neq S_z^B$). The AFM phase is stationary, which means that $S_z^{\alpha}$ will not change with time when $t\to +\infty$. In the OSC phase, however, populations of two neighboring sites oscillate over time.
	
	\subsection{Phase diagrams}
\begin{figure}
		\centering
		\includegraphics[width=1\linewidth]{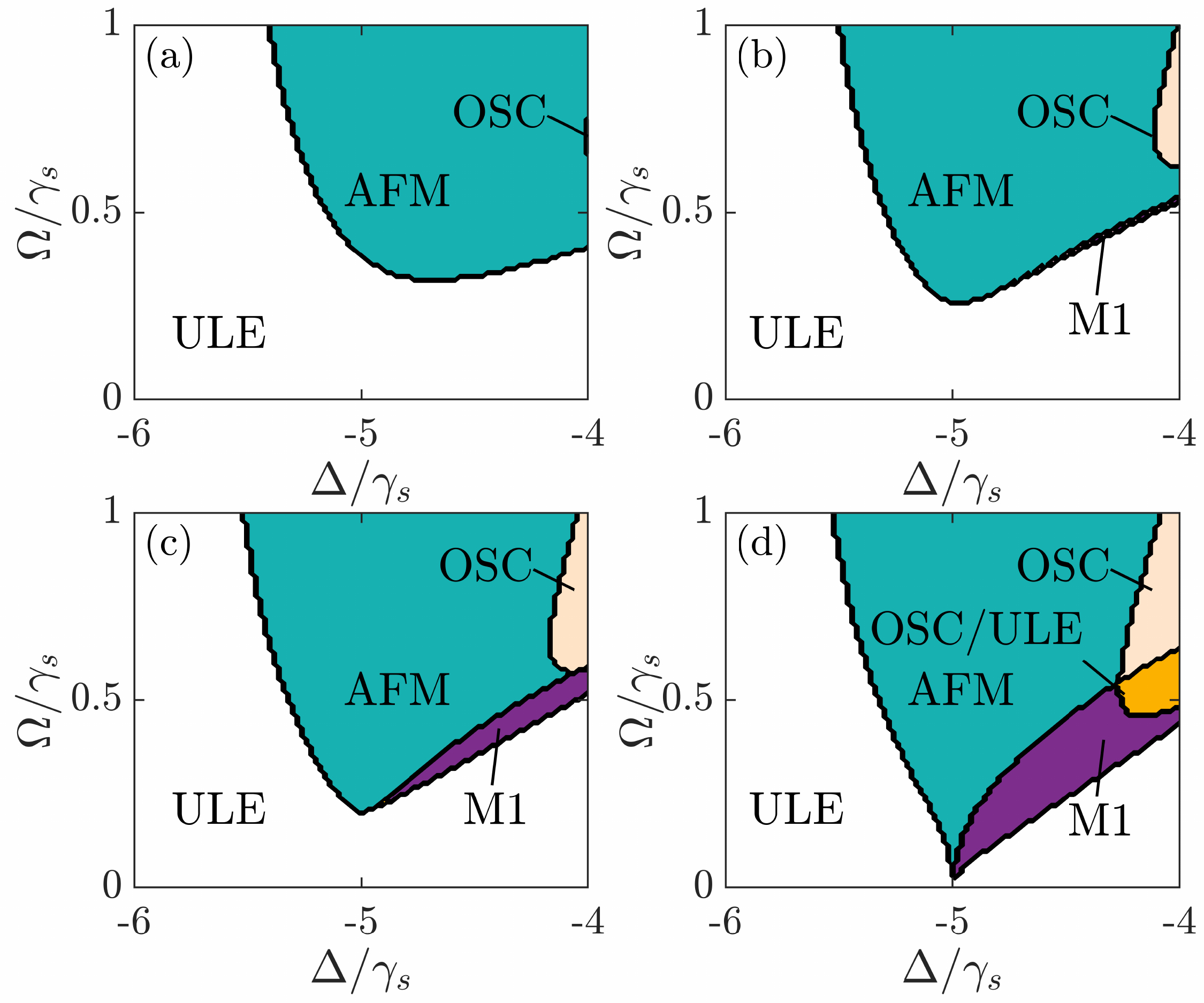}
		\caption{Mean-field phase diagrams for (a) $\gamma_m = 0$, (b) $\gamma_m = 0.6$, (c) $\gamma_m = 0.8$, and (d) $\gamma_m = 1$. When $\gamma_m$ is small, the phase diagram is mainly occupied by the UNI phase (ULE and UHE), AFM phase, and OSC phase. When $\gamma_m$ is large, bistable (OSC/ULE with orange area) and multistable phases (M1 with purple area) can be seen in panels (b), (c), and (d). Other parameters are $V_{1}=10$, $V_{2}=10$, and $\gamma_s = 1$.}
		\label{fig:N=2phase-gam}
	\end{figure}

	Examples of MF phase diagrams for different values of $\gamma_m$ are shown in Fig.~\ref{fig:N=2phase-gam}.
	They elaborate on the consequences arising from the nonlocal character of the dissipation. Without the nonlocal decay ($\gamma_m=0$) (Fig.~\ref{fig:N=2phase-gam}(a)), the steady state is dominated by a ULE phase when $\Omega$ is small and $\Delta<0$. By decreasing $|\Delta|$ and increasing $\Omega$, the ULE phase becomes unstable and enters into the AFM phase, due to the competition between the local decay and vdW interaction~\cite{Lee2014dissipative}. It is found that the OSC phase emerges when roughly $\Omega>0.6$ and $\Delta>-4$. More details on the phases without superradiance can be found in Appendix~\ref{No superradiance}. The presence of the Rydberg superradiance enhances the nonuniform phase and also brings multistable phases. As shown in Fig.~\ref{fig:N=2phase-gam}(b)-(c), areas of the ULE phase shrink when increasing $\gamma_m$, while areas of the nonuniform phase, especially the OSC phase, increase drastically. Importantly, a new \textit{multistable} phase (labeled by M1) emerges in which the AFM, OSC, and ULE phases coexist. For example, we find both a bistable region of the OSC and ULE phase, and a M1 phase when $\gamma_m=\gamma_s$, as depicted in Fig.~\ref{fig:N=2phase-gam}(d).

\begin{figure}
		\centering
		\includegraphics[width=1\linewidth]{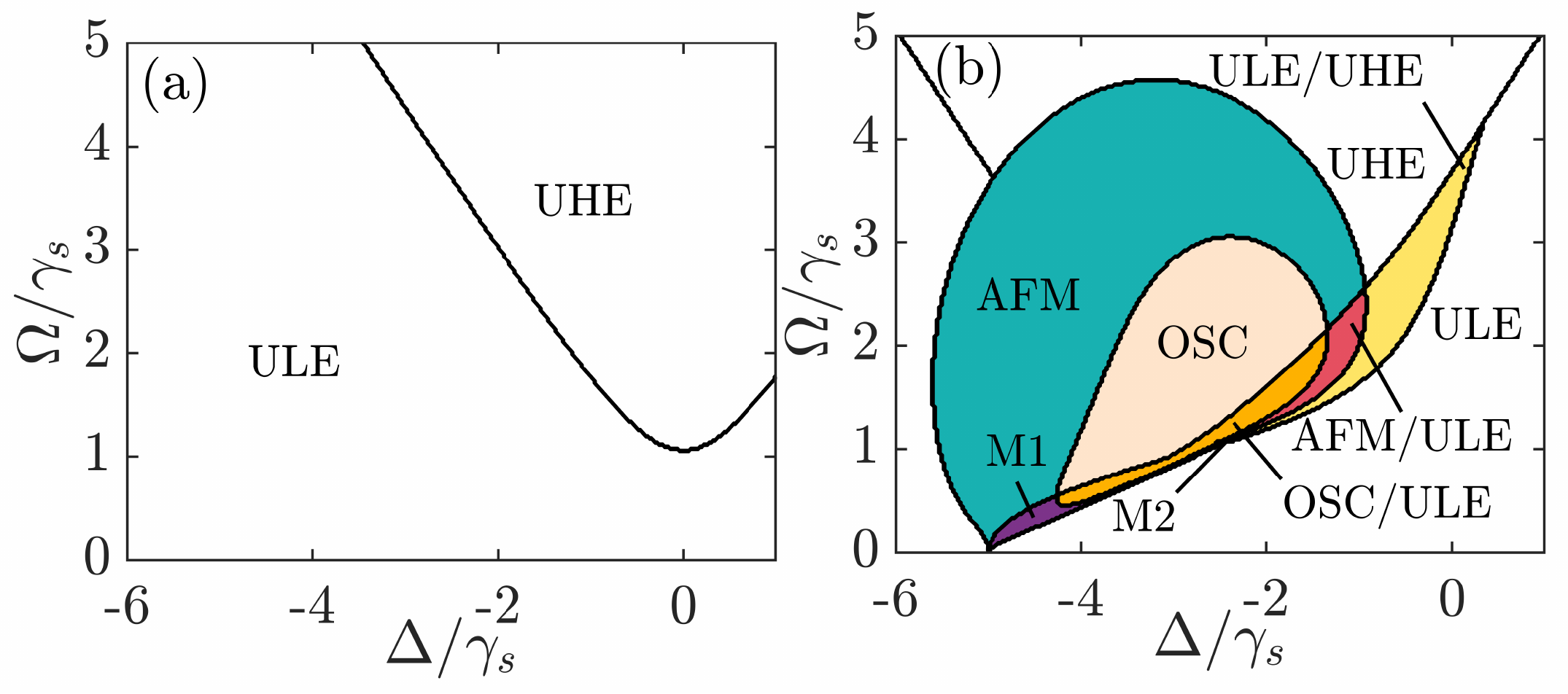}
		\caption{Homogeneous phases when (a) $V_{1}=V_{2}=0$ and more structured phases when (b) $V_{1}=V_{2}=10$. These figures show that the strong vdW interactions turn the simple uniform phase into complicated phases. Parameters are $\gamma_s=\gamma_m=\gamma=1$.}
		\label{fig:N=2phase-V}
	\end{figure}

	The rich MF phases result from the competition between the collective decay and strong vdW interaction. Without the vdW interaction, we only find uniform phases, as shown in Fig.~\ref{fig:N=2phase-V}(a). Here the ULE phase smoothly crosses over into the UHE phase as $\Omega$ is increased while $\Delta$ is fixed. When $V_{1} = V_{2} = 10$, on the other hand, a variety of nonuniform phases are generated, as shown in Fig.~\ref{fig:N=2phase-V}(b). Here even in the UNI phase, we find a bistability between the ULE/UHE phases when $|\Delta|\sim 0$. Note the bistable ULE/UHE phase is different from the AFM phase in that the population in the $A-B$ site is still same in the former case. The transition to these steady phases depends on initial conditions~\cite{lee2011antiferromagnetic, parmee2018phases}, which will be demonstrated in detail in the next section. Other bistable phases, including AFM/ULE and OSC/ULE phases, are also found, though they occupy a small parameter space. We also find a new multistable phase in which AFM(OSC)/ULE/UHE solutions (labeled by M2; see Fig.~\ref{fig:N=50multi}(a1) below for detail) are found. This multistable phase can only occupy a very small region in the parameter space. Hence the vdW interaction and nonlocal dissipation between different atoms together result to complicated phases~\cite{parmee2018phases}. In the following, we will focus on the bistable and the multistable M2 phases.

	\section{Stability and dynamics of the mean-field phases}\label{Analysis of mean field phases}
	The phase diagram obtained previously is based on mean-field calculations with Eq.~(\ref{eq:UNI-SZ}) (uniform phases) and Eq.~(\ref{eq:AB}) (bistable and multistable phases). In the following, we will study stabilities of these phases in a long chain $N \gg 2$, and hence verify especially the stability of the M1 phase.
	\subsection{Linear stability analysis}\label{Linear stability analysis}
	We first present examples of the multistability and bistability as a function of $\Omega$ in Fig.~\ref{fig:N=50multi}(a1). The blue lines represent the uniform solutions and the orange lines represent the nonuniform solutions. We then analyze the linear stability of the steady state solution by calculating eigenvalues $\lambda_j$ of the Jacobian matrix of Eqns. (\ref{eq:AB})~\cite{Strogatz2015Nonlinear}. If the real parts of all eigenvalues are negative, i.e. $\text{Re}(\lambda_j) < 0$, the corresponding solution is stable (solid lines); otherwise, it is unstable (dotted lines).
	
	When $\Omega$ is small, the steady state is the ULE phase, then changes to the OSC/ULE phase and then to the M1 phase by increasing $\Omega$ (Fig.~\ref{fig:N=50multi}(a1)). The nonuniform fix points become stable which means the system shows an antiferromagnetic pattern. These unstable nonuniform fixed points lead to the OSC phase, in which the Rydberg population oscillates periodically in time. In particular we find multistable solutions in the M1 phase (AFM/OSC/ULE), ULE solutions are stable while two other solutions are not stable. Further increasing $\Omega$, the nonuniform solutions become unstable while the UHE phase becomes stable at a critical $\Omega_c$ (marked by $C$) after passing through the very narrow M2 phase and the ULE/UHE phase.
	

	
	\subsection{Dynamics of the multistable phase}
\begin{figure}
		\centering
		\includegraphics[width=1\linewidth]{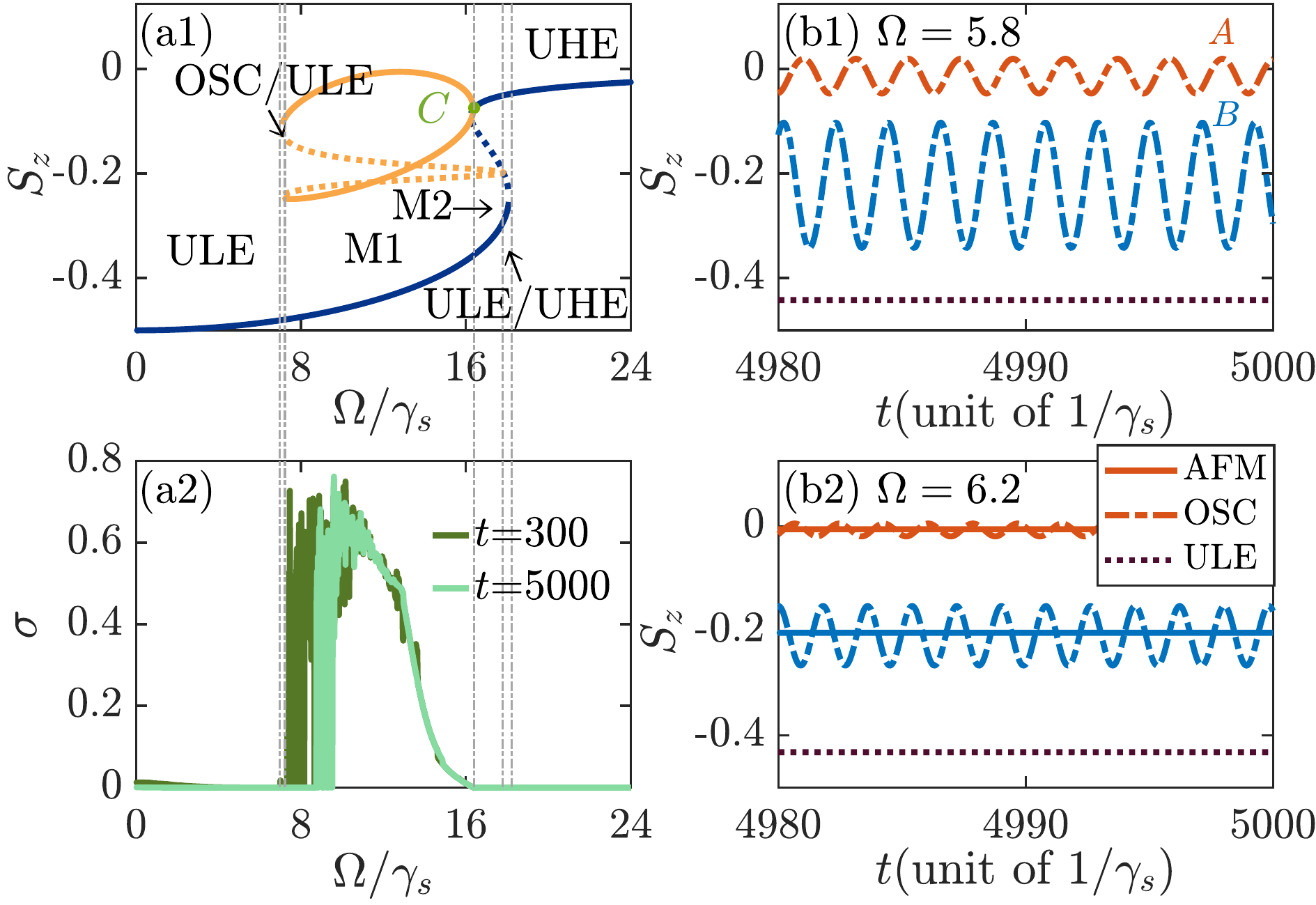}
		\caption{Superradiance dependent multistability. When increasing $\Omega$,  a series of phase transitions can be found in (a1). The solid (dotted) lines correspond to the stable (unstable) solutions. The blue (orange) lines correspond to the uniform (nonuniform) solutions. Multiple solutions coexist to represent the multistable phase. Point $C$ is the critical point between the nonuniform and UHE phase. The grey dash lines indicate the crossing into OSC/ULE, M1, M2, ULE/UHE, and UHE regions, respectively. Dynamics of Rydberg populations in the (b1) M1 phase with $\Omega = 5.8$ and (b2) M1 phase with $\Omega = 6.2$. Different curves represent different simulations. The red (blue) lines denote the dynamical behaviors of atoms in site $A$ ($B$) of the simulation. The solid lines correspond to the AFM phase, the dash lines correspond to the OSC phase, and the dotted lines correspond to the ULE phase. (a2) Variance $\sigma$ as a function of Rabi frequency in the long-time limit $t=300$ and $t=5000$. Other parameters are $N = 50$, $\Delta = -2$, $V_{1} = V_{2} = 10$, and $\gamma_s = \gamma_m = 0.5$. }
		\label{fig:N=50multi}
	\end{figure}

	In the multistable phase, atoms at different sites can occupy different stable populations. To verify this, we solve Eq.~(\ref{eq:MF}) numerically with $N=50$ and periodic boundary conditions. The initial values of different atoms are $\{S^i_x, S^i_y, S^i_z\}=\{0,0,r\}$, where $r$ is a random number between $-0.5$ and $0.5$. We then probe the multistable phase by tuning the parameters. In Fig.~\ref{fig:N=50multi}(b1) and (b2), we show mean values of $S_z^j$ for a block of $6$ sites with index $j=1\sim6$.  As shown in Fig.~\ref{fig:N=50multi}(a1), the simple two-site MF theory predicts three stable solutions in the M1 phase, which can be seen in the dynamical simulation with $N=50$. We note that in the many site simulation, the system prefers a ULE and OSC solution when $\Omega$ is approaching to the lower critical value around $5.8$, as the example shown in Fig.~\ref{fig:N=50multi}(b1). Increasing $\Omega$, the three phases coexist in the dynamical simulation, as shown in Fig.~\ref{fig:N=50multi}(b2). The OSC phase oscillates around the AFM phase and its oscillation amplitude reduces with the Rabi frequency. Further increasing $\Omega$, the strength of the OSC phase gradually reduces such that only the AFM and ULE phase survive.
	
	To characterize distributions of the Rydberg spin population across the lattice, we evaluate the variance $\sigma$ of the spins in different sites~\cite{parmee2018phases}
	\begin{equation}\label{eq7}
		\sigma=\frac{1}{N}\sum_i^{N}(\bar{\mathbf{S}}-\mathbf{S}^i)^2,
	\end{equation}
	where $\mathbf{S}^i=(S_x^i,S_y^i,S_z^i)/S, S=\sqrt{(S_x^i)^2+(S_y^i)^2+(S_z^i)^2}$, and $\bar{\mathbf{S}}=\sum_j^{N}\mathbf{S}^j/N$  is the average spin. Here the translational symmetry of the lattice is broken when $\sigma \neq 0$, which takes place, for example, in the AFM phase~\cite{Lee2014dissipative}. In Fig.~\ref{fig:N=50multi}(a2), we show the variance obtained from a simulation by varying $\Omega$. The spin fluctuations are large especially in the M1 phase due to different sites occupying very different populations. In the M1 phase, we find the variance reaches maximal values when the OSC phase dominates. It decreases when increasing $\Omega$, as the strength of the OSC phase decreases, while the AFM and ULE phase become important. We have evaluate the values at two different times. It is found that the spin fluctuation persists even when $t=5000$, indicating that the various phases are truly stable. Note that in the bistable phases, the atoms will pick up either the lower or the upper branch of the solution in individual simulations, hence $\sigma=0$ in these phases.

	\subsection{The critical value $\Omega_c$ }
\begin{figure}
		\centering
		\includegraphics[width=1\linewidth]{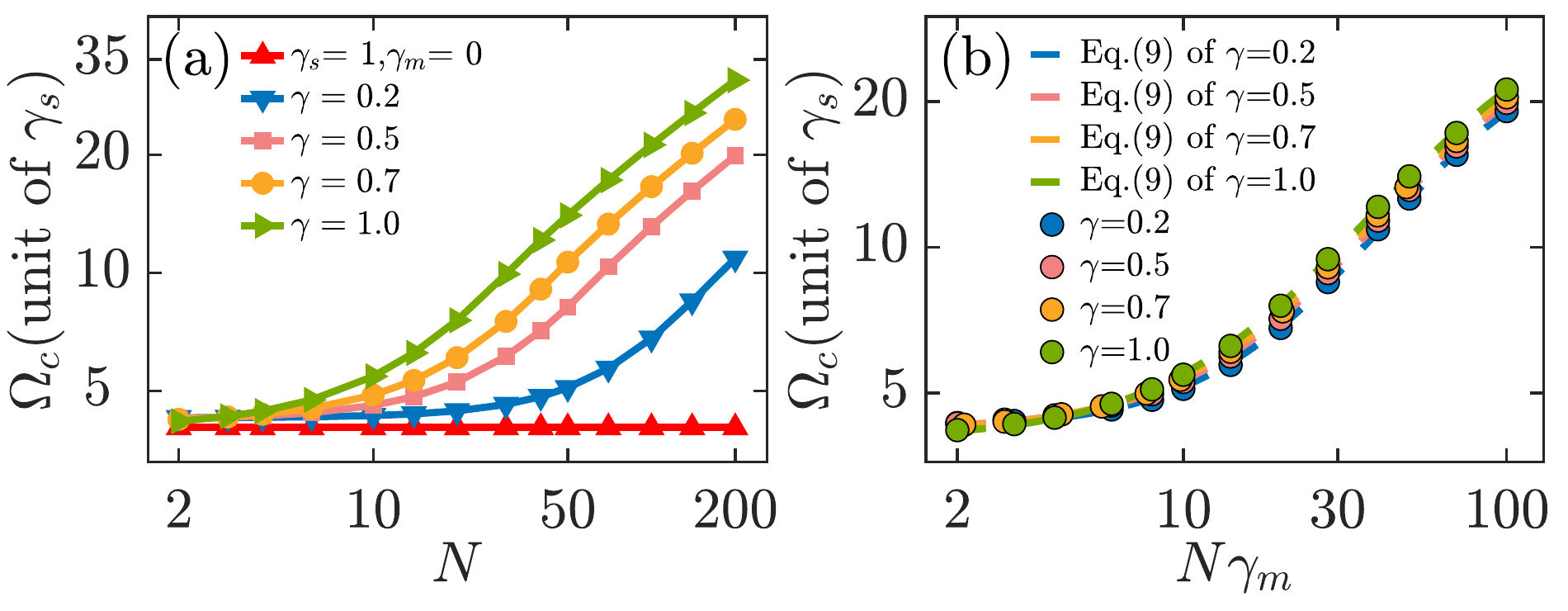}
		\caption{Scaling of the critical value $\Omega_c$ with respect to (a) atom numbers $N$ and (b) parameter $N\gamma_m$ for different $\gamma$. (b) The colored dots denote numerical simulation, and the lines represent analytical results from Eq.~(\ref{eq:critical}).}
		\label{fig:scaling}
	\end{figure}

	As shown in Fig.~\ref{fig:N=50multi}(a1), point $C$ marks the boundary between the M1 and UHE phase.  It is interesting to understand the critical value $\Omega_c$ that distinguishes these two phases. When increasing $\gamma_s = \gamma_m = \gamma$, our numerical simulations indicate that  $\Omega_c$ increases, as shown in Fig.~\ref{fig:scaling}(a). In addition, the critical value increases with $N$ monotonically for a given $\gamma$, as the effective collective decay rate of each atom is proportional to $(N-1)\gamma_m$.  Note that $\Omega_c$ can only be tuned in the superradiance regime. When $\gamma_m=0$, it will be a constant and has no dependence on $N$ any more.
	
	
	As shown in Fig.~\ref{fig:scaling}(a), $\Omega_c$ increases nearly linearly when $N$ and $\gamma_m$ are large, which displays different scaling when $N$ and $\gamma_m$ are small. To understand this behavior, one notes that the critical point can be obtained by solving Eq.~(\ref{eq:AB}). Approaching the critical point from the UHE phase, $S_z$ is solved numerically using Eq.~(\ref{eq:UNI-SZ}). We derive an analytical solution,
	\begin{equation}\label{eq:critical}
		\Omega_c=\sqrt{-\frac{2S_z+1}{S_z}\left[(\Delta+S_{V})^2\!+\!\left(\frac{\gamma_s}{2}-(N-1)\gamma_m S_z\right)^2\right]}.
	\end{equation}
	The analytical $\Omega_c$ shows that the critical point will depend on $N$ if $\gamma_m\neq 0$. When $N\gamma_m<\gamma_s$, $\Omega_c$ varies with $S_V$ and $\gamma_s$ nonlinearly. Hence this is a regime where the vdW interaction dominates, as $S_V$ is affected by the vdW interaction. When $N\gamma_m$ is large, on the other hand, one can expand $\Omega_c$ by assuming $\gamma_s$ and $S_V$ small, leading to $\Omega_c\sim \sqrt{-S_z(2S_z+1)} (N-1)\gamma_m$. In Fig.~\ref{fig:scaling}(b), the scaled $\Omega_c$ with respect to $N\gamma_m$ are shown. The numerical data agree with the analytical prediction $\Omega_c$ well.

	\section{Quantum many-body dynamics of finite 1D chains}\label{Quantum many-body dynamics of finite 1D chains}
	
	\begin{figure}
		\centering
		\includegraphics[width=1.0\linewidth]{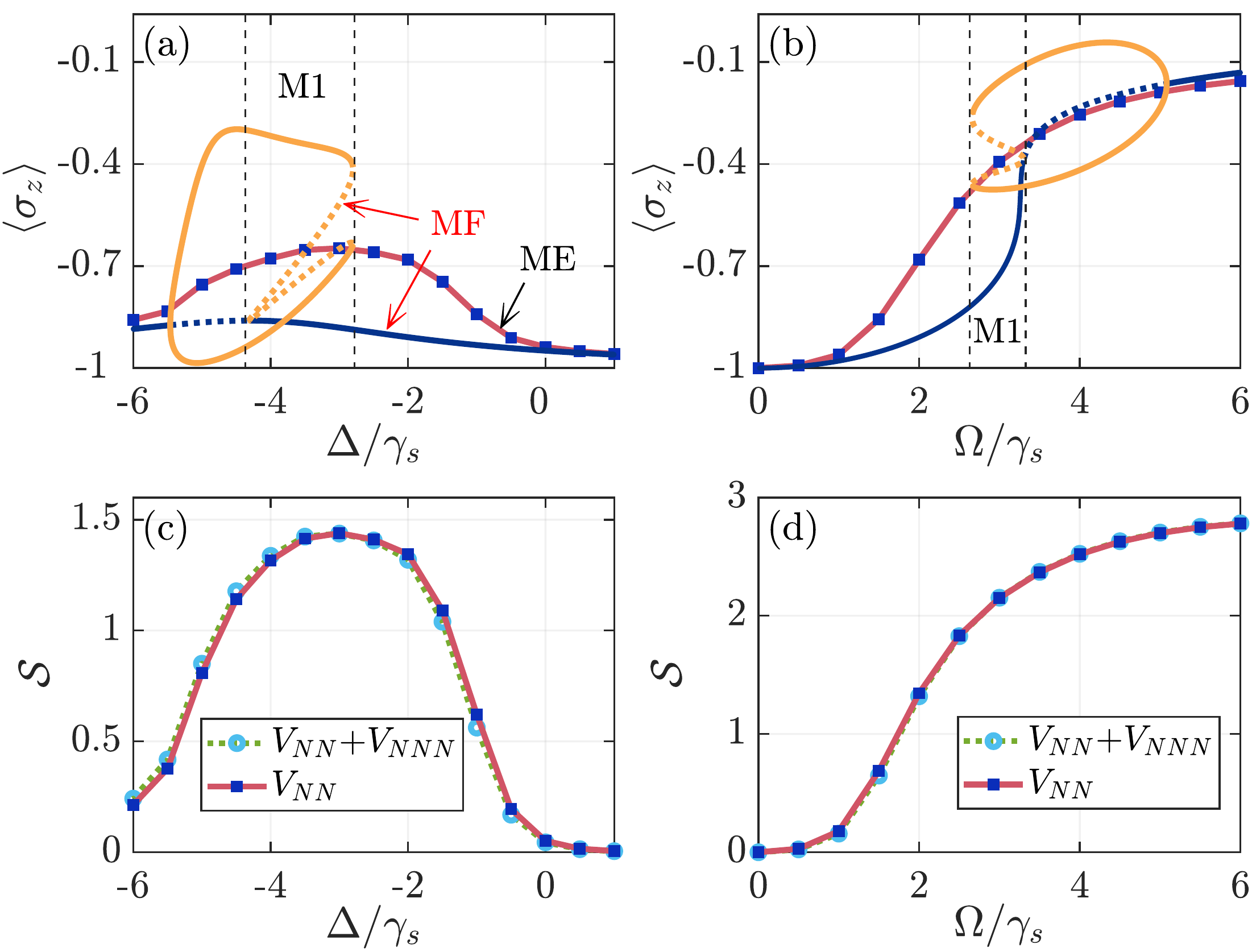}
		\caption{Numerical solutions of the master equation (ME) and MF calculations for 1D chain of length $N = 8$ with periodic boundary conditions. Mean population $\langle \sigma_z\rangle$ by varying (a) detuning $\Delta$ with $\Omega = 2$ and (b) Rabi frequency $\Omega$ for $\Delta = -2$. The red lines represent the master equation results. The blue and orange curves are MF results. The M1 phase is highlighted in (a) and (b). The tendency of the MF and master equation calculation is similar. It seems that the M1 phase emerges before $\langle\sigma_z\rangle$ reaches the maximal value when increasing $\Delta$ or $\Omega$.  The respective Von Neumann entropy in panel (c) and (d) has similar shapes as that of $\langle \sigma_z\rangle$ shown in (a) and (b). The dark blue squares are calculated by only considering the NN interaction ($V_{NN}$) while the light blue circles by considering both NN and the next NN interaction ($V_{NN}$ and $V_{NNN}$). Other parameters are $N=8$, $\gamma = 1$, $V_{1} = 5$, and $V_{2} = 5$.}
		\label{fig:S<Sz>}
	\end{figure}
	
	MF theory is expected to be valid in higher dimensions where quantum fluctuations are averaged out. Despite this, MF theory can capture qualitative aspects of the quantum system. To illustrate signatures of the MF phases, we numerically solve the master equation~(\ref{eq:master}) for a 1D chain of length $N = 8$ with periodic boundary conditions in the long-time limit $t=300$. In Fig.~\ref{fig:S<Sz>}(a) and (b), mean values of spin population, $\langle \sigma_z\rangle= 1/N\sum_j\text{Tr}(\rho_s\sigma_z^j)$, in the stationary state $\rho_s$ are shown. It is found that some trends of the master equation calculation agree with the MF prediction. For example, in the M1 (Fig.~\ref{fig:S<Sz>}(a) and (b)), mean values of the spin component $\sigma_z$ becomes large when varying $\Delta$ or $\Omega$. This means spin state $|2\rangle$ is excited in these parameter regions. A consequence is that the von Neumann entropy $\mathcal{S} = -\text{Tr}(\rho_s\ln \rho_s)$ in the steady state also becomes large (Fig.~\ref{fig:S<Sz>}(c) and (d)). As shown in Fig.~\ref{fig:S<Sz>}(c) and (d), even longer range interactions (i.e. next nearest-neighbor interactions) only plays a minor role, justifying that it is a good approximation to consider only the nearest-neighbor interaction in the calculation.
	
\begin{figure}
		\centering
		\includegraphics[width=1\linewidth]{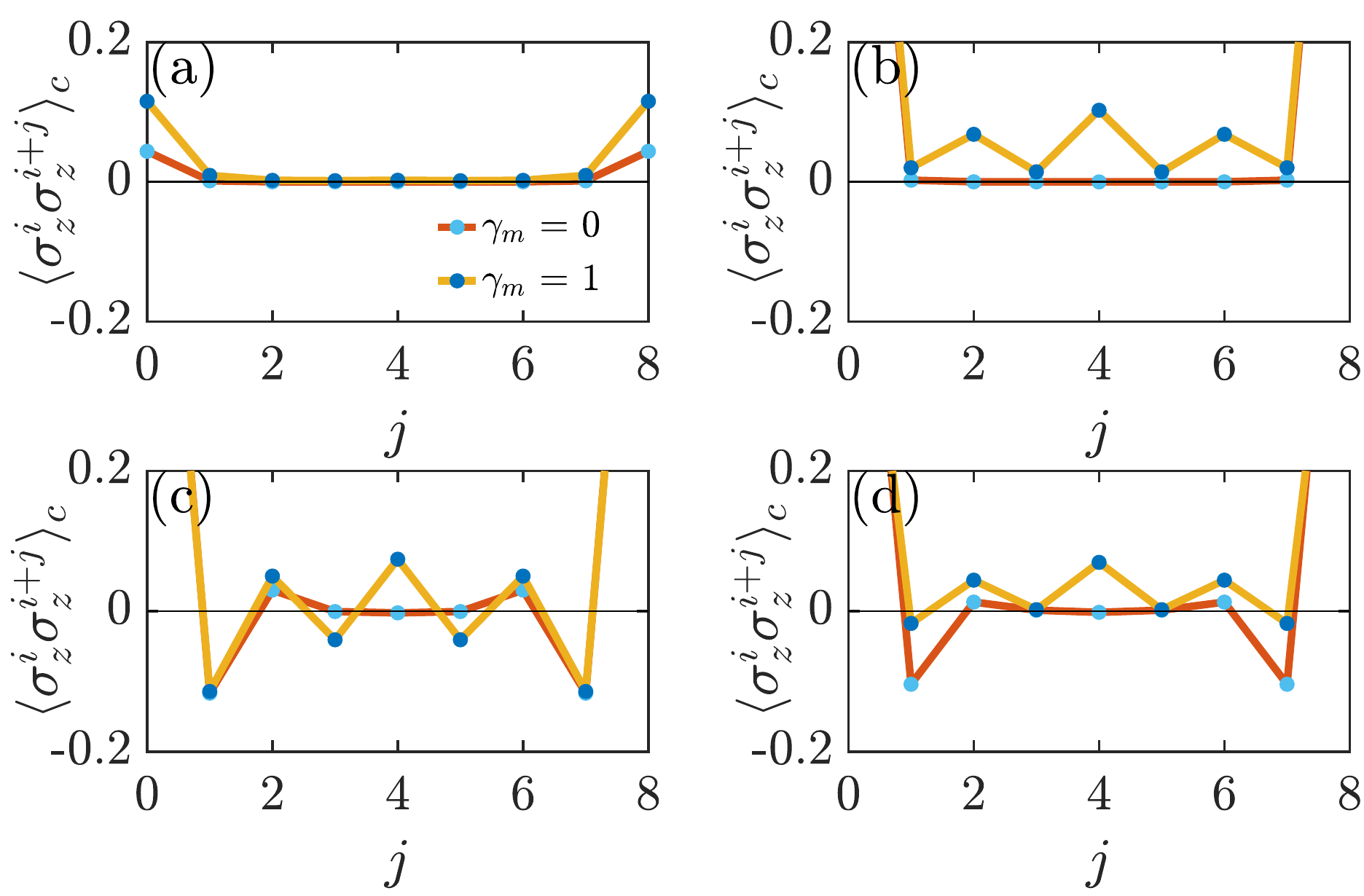}
		\caption{The correlation operator $\langle \sigma_z^i \sigma_z^{i+j}\rangle_c$ as a function of distance with $\gamma_s = 1$, $V_1 = V_2 = 5$ for $N = 8$ spins on a chain with periodic boundary conditions. (a) ULE phase $\Delta=0$, $\Omega=2$, (b) UHE phase $\Delta=0$, $\Omega=8$, (c) AFM phase $\Delta=-3$, $\Omega=4$, (d) OSC (M1) phase $\Delta=-3, \Omega=2.4$ for $\gamma_m=0(1)$.}
		\label{fig:correlation}
	\end{figure}

	Another important quantity is the correlation between different lattice sites, $\langle \sigma_z^i \sigma_z^{i+j}\rangle_c=\langle \sigma_z^i \sigma_z^{i+j}\rangle-\langle \sigma_z^i\rangle\langle \sigma_z^{i+j}\rangle$~\cite{Lee2014dissipative}.  Due to a periodic boundary condition, the correlation will vary with the lattice separation. For concreteness, we consider $i=1$ and $0\le j \le 8$ in the calculation. The correlation exhibits rather different features in different MF phases.
	In the ULE phase, the correlation decays rapidly with increasing distance and vanishes when $j>1$, which is independent of $\gamma_m$, shown in Fig.~\ref{fig:correlation}(a). In the ULE phase, atoms in the system are largely in the low-lying $|1\rangle$ state. Hence jumping from state $|2\rangle$ to $|1\rangle$ is unlikely, such that the stationary state as well as the correlation is largely insensitive to $\gamma_m$. This, however, changes in the UHE phase, where the occupation in state $|2\rangle$ in every site is large. In this phase, the superradiance plays an important role in the stationary state. As shown in Fig.~\ref{fig:correlation}(b), a long-range, positive correlation is obtained when $\gamma_m=1$, while the correlation does not exist any more when $\gamma_m=0$. In the AFM phase (Fig.~\ref{fig:correlation}(c)), we find that the correlation oscillates between positive and negative values with increasing $j$ when $\gamma_m=1$. In the M1 phase region, however, the correlation is negative when $j=1,7$ , and becomes positive at large separations (Fig.~\ref{fig:correlation}(d)). The correlation, however, decays with increasing separation when $\gamma_m=0$. This indicates that the nonlocal decay can enhance the two-body correlation. Hence the different profiles of the spin-spin correlation could be used to characterize the MF phases.
	
	
	\section{Conclusions}\label{discussion and conclusions}
	We have investigated stationary phases of a 1D chain of MW coupled, strongly interacting Rydberg atoms with nonlocal dissipations. Using MF theory, we have obtained interesting  bistable and  multistable solutions in the stationary state. By analyzing the MF phase diagram, the dependence of the multistable phases on the MW coupling, nonlocal dissipation as well as vdW interaction is studied. Dynamical simulations show that Rydberg atoms in different sites  occupy all available solutions simultaneously in the multistable phase. We have found the critical value $\Omega_c$ between the multistable and UHE phase. The scaling of $\Omega_c$ is examined, and agrees with numerical calculations. By solving the master equation numerically for a finite chain, it is found that certain features predicted by the MF theory persist in the quantum regime. Different profiles of the spin-spin correlation could be used to probe and characterize the MF phases. Such superradiance induced many-body phase transition is observable with current experimental condition\cite{orioli2018relaxation, hao2021observation}. Our study is relevant to current theoretical~\cite{nill_many-body_2022} and experimental~\cite{hao2021observation} efforts in understanding and probing dynamics due to the interplay between strong vdW interactions and superradiant decay in arrays of Rydberg atoms.

	
	\begin{acknowledgments}
		Y. H., Y. J., and J. Z. are supported by the National Natural Science Foundation of China (Grant No. 12120101004, 61835007, 62175136); the Scientific Cooperation Exchanges Project of Shanxi province (Grant No. 202104041101015); Changjiang Scholars and Innovative Research Team in Universities of the Ministry of Education of China (IRT 17R70); the Fund for Shanxi 1331 Project. Z. B. acknowledge National Natural Science Foundation of China (11904104, 12274131), and the Shanghai Pujiang Program under grant No. 21PJ1402500. W. L. acknowledges support from the EPSRC through Grant No. EP/W015641/1.
	\end{acknowledgments}

	\appendix
	\section{Dipole-dipole and vdW interactions}\label{experimental parameter}
	\setcounter{figure}{0}
	\renewcommand\thefigure{A\arabic{figure}}
	\begin{figure}[htp!]
		\centering
		\includegraphics[width=0.95\linewidth]{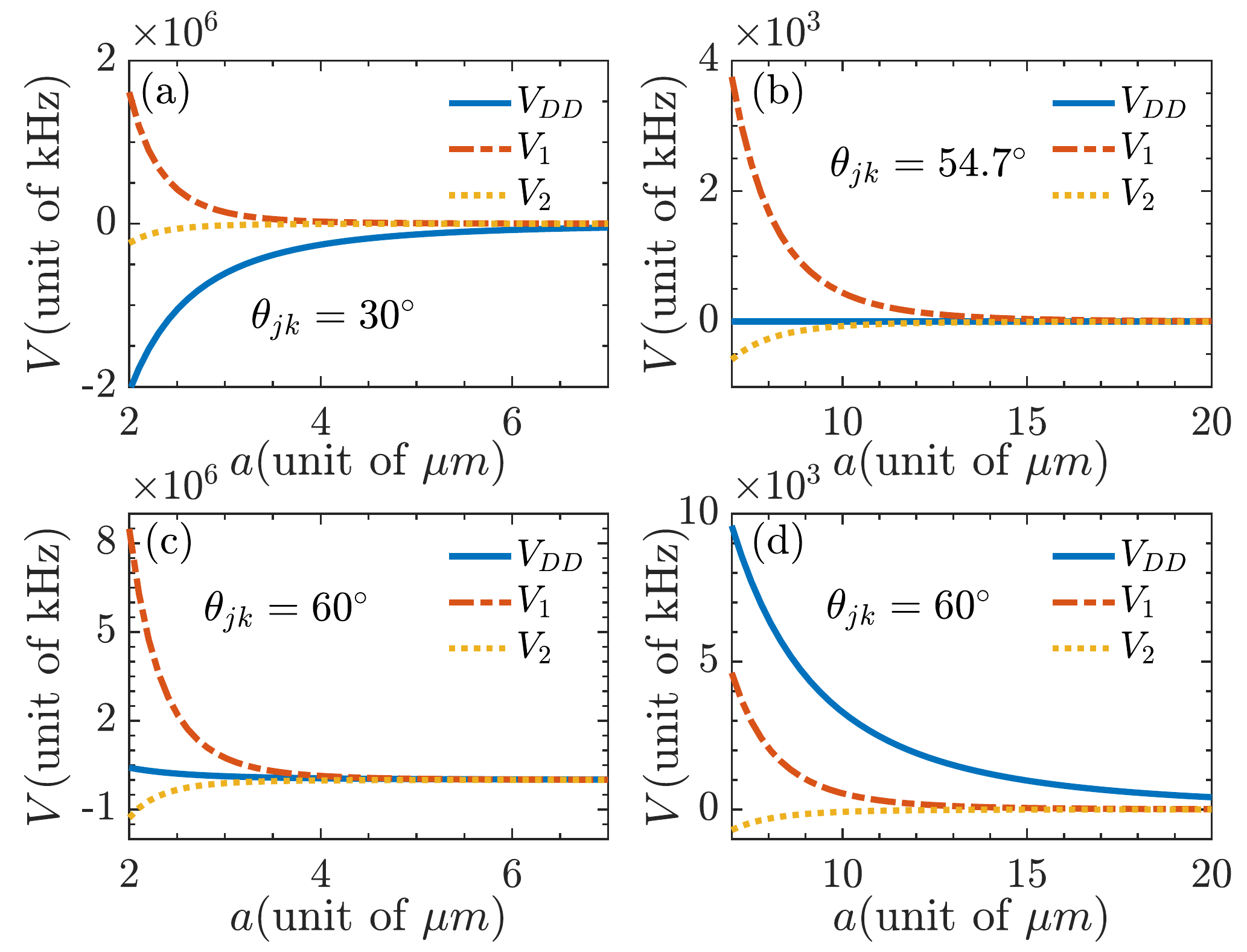}
		\caption{DD and vdW interactions. In (a) $\theta_{jk}=30^\circ$, where the vdW interaction is comparable to the DD interaction. In (b) $\theta=54.7^\circ$, the DD interaction is zero. In (c) and (d), $\theta_{jk}=60^\circ$. We can see that the vdW interaction is important at short distances (c), while the DD dominates when atom separations are large (d).}
		\label{DDvdW}
	\end{figure}
	In this section, we discuss the strength of both DD and vdW interactions and the motivation of neglecting the DD interaction in this work. The experimental and numerical results in our recent work~\cite{hao2021observation} demonstrate that the dipolar interaction effect might not be critical in dense gas. This is because that dipolar interactions are a long-range interaction ($\sim R^{-3}$), but the van der Waals interaction is short-ranged ($\sim R^{-6}$). For high atomic density, the distance between atoms is small ($R \le \lambda$), where the vdW interaction could play a dominant role (see Fig.~\ref{DDvdW} for illustrations). Moreover, we can control the strength of the DD interaction by manipulating the angle $\theta_{jk}$.  To highlight the contribution of vdW interaction in one-dimensional system, we can adjust the magic angle ($1-3\cos^2\theta_{jk}=0$) to turn off the DD interaction. Hence the DD interactions can be safely ignored in our model (see blue line in Fig.~\ref{DD_vdW_Compare}). 
	\begin{figure}[htp!]
		\centering   \includegraphics[width=1\linewidth]{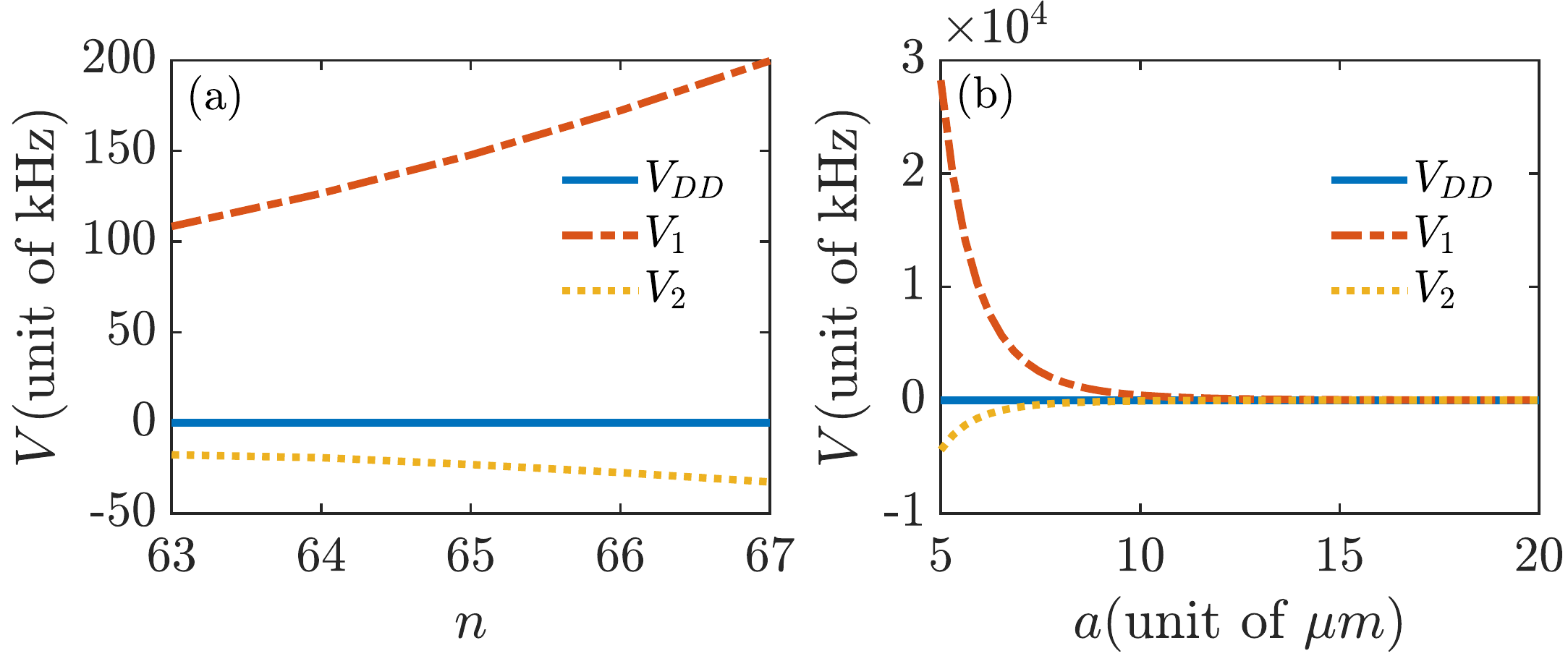}
		\caption{(a) The interaction energy (including DD and vdW interactions) as a function of the principal quantum number $n$ with lattice constant $a=12\mu m$. (b) The interaction energy varies with lattice spacing $a$ when $n=65$. Here we select the magic angle where $\theta_{jk}=\arccos(\frac{1}{\sqrt{3}})\approx54.7^\circ$,  the dipole-dipole interaction $V_{DD}$ is close to zero.}
		\label{DD_vdW_Compare}
	\end{figure}
	\begin{figure}[htp!]
		\centering   \includegraphics[width=1\linewidth]{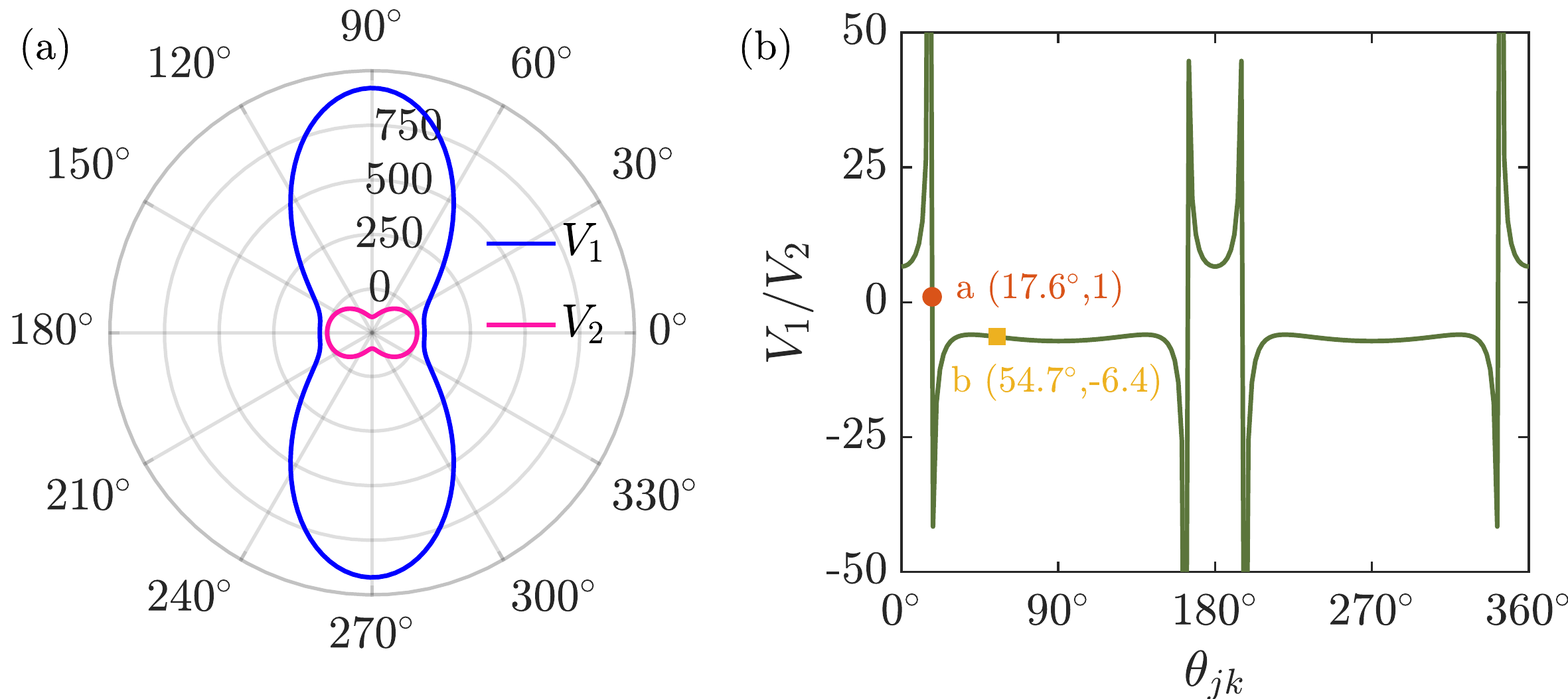}
		\caption{The calculation results with experimental parameters. (a) The dispersion coefficient $C_6$ varies with angular $\theta_{jk}$ . (b) The ratio between $V_1$ and $V_2$ as a function of $\theta_{jk}$. The point $a$ and $b$ indicates $V_1=V_2$ [the corresponding MF phase diagram is shown in Fig.~\ref{fig:N=2phase-V}(b)] and $V_1=-6.4V_2$ [the corresponding MF phase diagram is shown in Fig.~\ref{Phasediagram}], respectively.}
		\label{vdW_V1_V2_theta}
	\end{figure}
	
\begin{figure}[htp!]
	\centering   \includegraphics[width=0.8\linewidth]{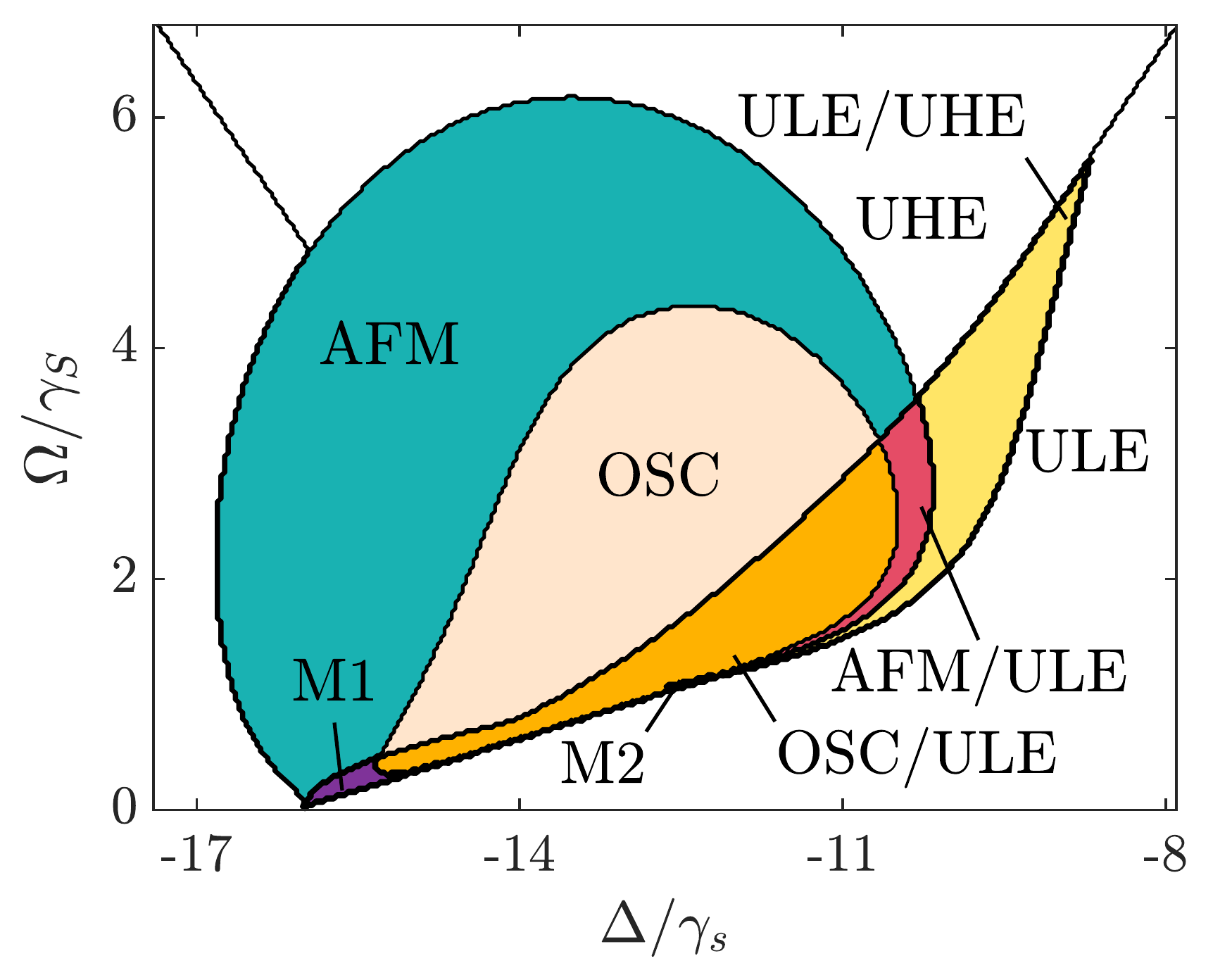}
	\caption{MF phase diagram with $V_{1}=32$ and $V_{2}=-5$. The superradiance rate is used with $\gamma_s=\gamma_m=1$.}
	\label{Phasediagram}
\end{figure}

	In this work cesium atoms are used with $|1\rangle=|(n+1)P_{3/2}\rangle$ and $|2\rangle=|nD_{5/2}\rangle$. The dispersion coefficient $C_6$ can be calculated using ARC package~\cite{ARC_2021}. The results shows that dispersion coefficients in states $|(n+1)P_{3/2}\rangle$ and $|nD_{5/2}\rangle$  are  both  anisotropic [see Fig.~\ref{vdW_V1_V2_theta}(a)]. From Fig.~\ref{vdW_V1_V2_theta}(b), one can see that the ratio between $V_1$ and $V_2$ can be precisely controlled by manipulating the angle $\theta_{jk}$. The condition for $V_1=V_2$ is achievable in our system when $\theta_{jk}\approx17.6^\circ$ [see point $a$ in Fig.~\ref{vdW_V1_V2_theta}(b)]. We also conduct the  simulation at the magical angle with interaction strength $V_1=32$, $V_2=-5$ [correspond to point $b$ in Fig.~\ref{vdW_V1_V2_theta}(b)]. Their corresponding MF phase diagram is shown in Fig.~\ref{Phasediagram}. Similar to the result given in Fig.~\ref{fig:N=2phase-V}(b) in the main text, abundant many-body phases can also be obtained here.
	

\section{Analytical Solutions of the uniform phase}\label{Uniform Analytical Solutions}

When $V_1=-V_2=V$, $\tilde{\Delta}=\Delta+V/2$, we can obtain the uniform solutions analytically,
\begin{equation}
	S_z=\frac{1}{12\kappa}\left\{-2\kappa+4\gamma_s+\frac{4(\Gamma^2-6c_1)}{(\sqrt{3}i-1)c_3^{\frac{1}{3}}}+(\sqrt{3}i-1)c_3^{\frac{1}{3}}\right\}
\end{equation}
where we have defined parameters,
\begin{subequations}
	\begin{align}
		\Gamma=&\kappa+\gamma_s,\nonumber \\
		c_1=&2\tilde{\Delta}^2+\Omega^2,\nonumber\\
		c_2=&(-\Gamma^2+6c_1)^3 + [\Gamma^3 + 36\Gamma\tilde{\Delta}^2-9(\kappa-2\gamma_s)\Omega^2]^2,\nonumber\\
		c_3=&-{\kappa}^3-3\gamma_s\kappa^2-3[\gamma_s^2+3(4\tilde{\Delta}^2-\Omega^2)]\kappa-\gamma_s(\gamma_s^2\nonumber\\ &+18c_1)+\sqrt{c_2}.\nonumber
	\end{align}
\end{subequations}
This expression is lengthy and therefore is not shown in the main text. It agrees with the numerical simulation.

\section{MF Phases without Superradiance}\label{No superradiance}
\begin{figure}
	\centering
	\includegraphics[width=1\linewidth]{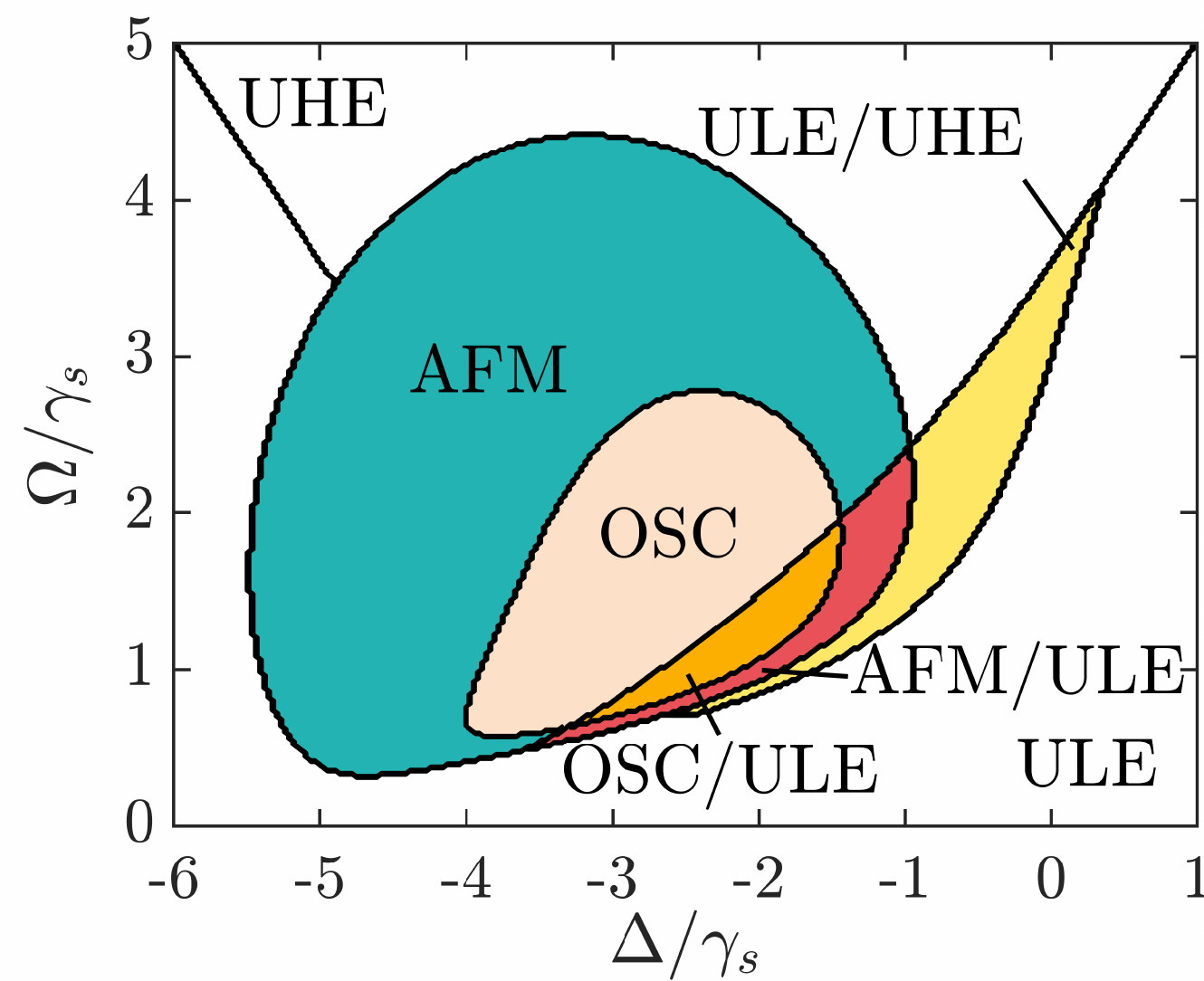}
	\caption{Mean-field phase diagram without superradiance. Other parameters are $V_{1}=10$ and $V_{2}=10$. }
	\label{fig:N=2nosuper}
\end{figure}
The mean-field phase diagram without superradiance is shown in Fig.~\ref{fig:N=2nosuper}. Similar work has been studied in Ref.~\cite{lee2011antiferromagnetic}. The difference is that the two-level system in our model consists of two Rydberg states. Compared to the superradiance phase diagram (see Fig.~\ref{fig:N=2phase-V}(b) in the main text), the influence of superradiance is negligible when the MW field driving is strong. Around $-5<\Delta<-1$, superradiance makes obvious changes. For example, the stable ULE phase in Fig.~\ref{fig:N=2nosuper} becomes nonuniform and emerges M1 phase.
\begin{figure}
	\centering
	\includegraphics[width=0.8\linewidth]{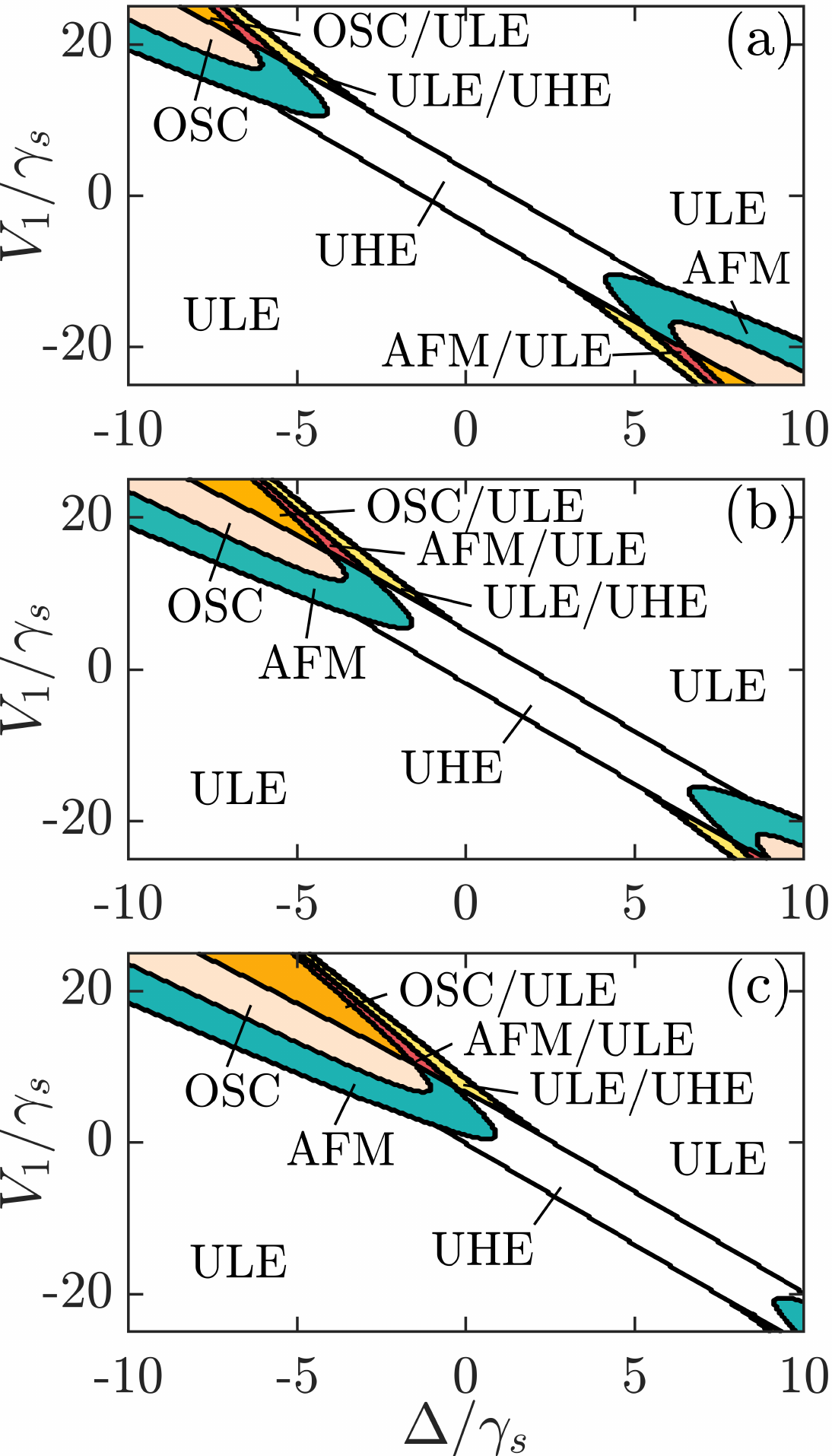}
	\caption{Mean-field phase diagrams for different $V_2$.  (a) $V_{2}=0$, (b) $V_{2}=5$, (c) $V_{2}=10$. The interaction $V_{2}$ acts as a detuning shift in the phase space. Here $\Omega=2$.}
	\label{fig:N=2no-interaction}
\end{figure}
\begin{figure}
	\centering
	\includegraphics[width=1\linewidth]{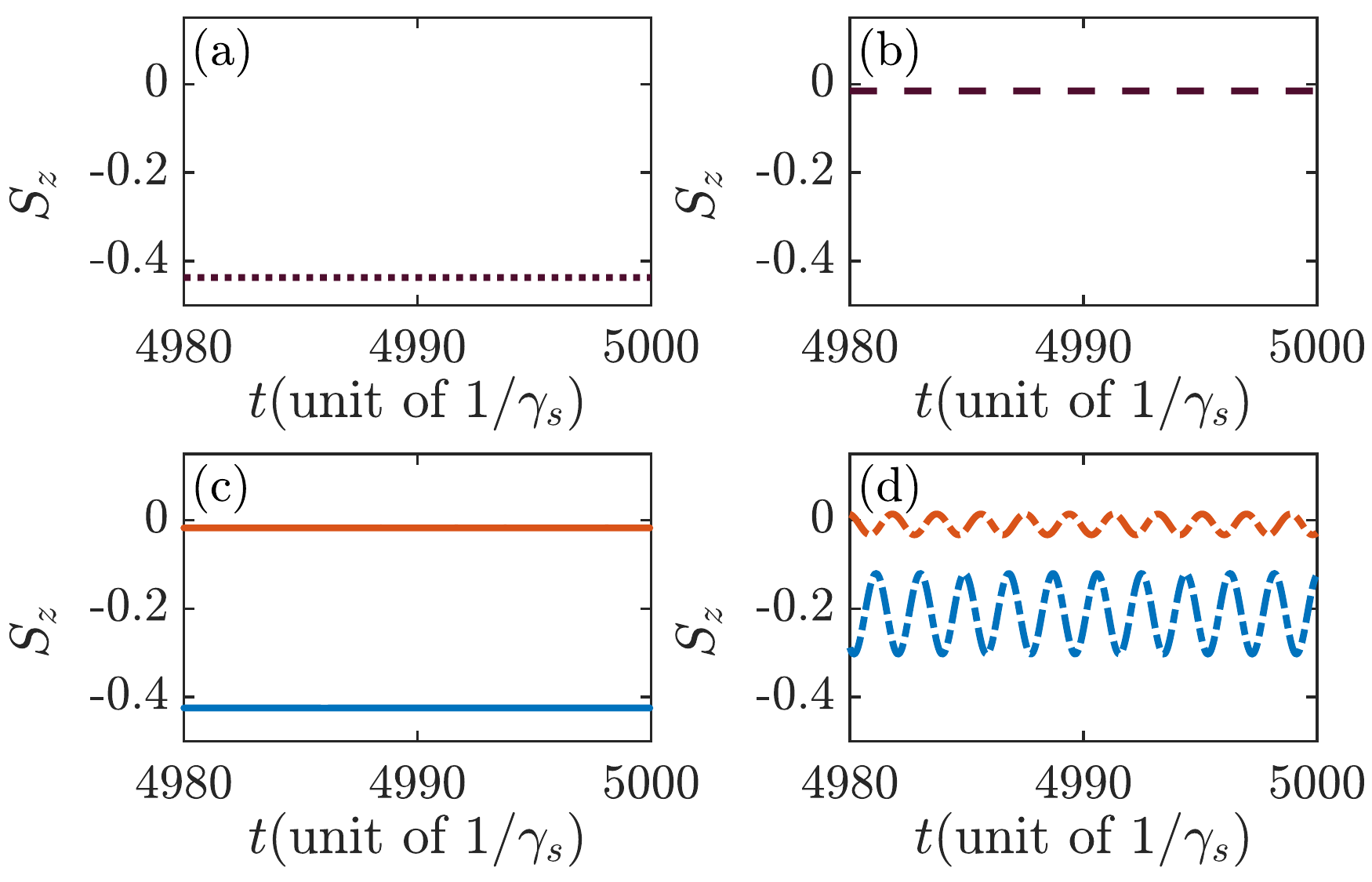}
	\caption{Dynamics of the order parameter $S_z^j$ ($j=1\sim6$) with different interaction and $N=50$, $\Delta=-2$, $\Omega=6$, $\gamma = 0.5$, (a) ULE phase with $V_{1}=0$, $V_{2}=5$, (b) UHE phase with $V_{1}=0$, $V_{2}=-5$, (c) AFM phase with $V_{1}=2.5$, $V_{2}=5$, (d) OSC phase with $V_{1}=5$, $V_{2}=5$. The purple dotted lines correspond to the UNI phase. The red (blue) lines denote the dynamical behaviors of atoms in site A (B) at the same simulation. The solid lines correspond to the AFM phase, the dash lines correspond to the OSC phase.}
	\label{fig:dynamicsN=50interaction}
\end{figure}

We further study the influence of vdW interaction on phase transitions. Fig.~\ref{fig:N=2no-interaction} shows the mean-field phase diagrams as a function of $\Delta$ and $V_{1}$ for $\Omega=2$ with the vdW interaction $V_{2}$ is equal to $0$, $5$, $10$, respectively. Fig.~\ref{fig:N=2no-interaction}(a) shows the phase is symmetric with respect to the origin, i. e. one always observes an identical phase at points ($\Delta$,$V_1$) and (-$\Delta$, -$V_1$). The central region of the phase diagram is affected by the driving field, i.e. sufficiently strong driving strength changes the system to the UHE phase. When the vdW interaction of the $|1\rangle$ state is weak, $|V_{1}| < 10$, there is only the UNI phase when scanning the detuning. As the detuning increases, a continuous phase transition occurs from the UHE phase to the ULE phase which means the atoms from the high-lying Rydberg state return to the lower-lying Rydberg state. For positive detuning and negative interaction $V_{1}$ which appears as an attractive potential, the uniform phase disappears, and the AFM phase emerges. With the further increase of the parameters, the AFM phase becomes unstable and develops into the OSC phase. A series of continuous phase transitions occur as the interaction $V_{2}$ increases. The increase of $V_{2}$ breaks the symmetry of the phase diagram and the symmetry point moves downward. The regions of the five phases  except the UNI phase increase at positive interaction $V_{1}$.

\begin{figure}
	\centering
	\includegraphics[width=1\linewidth]{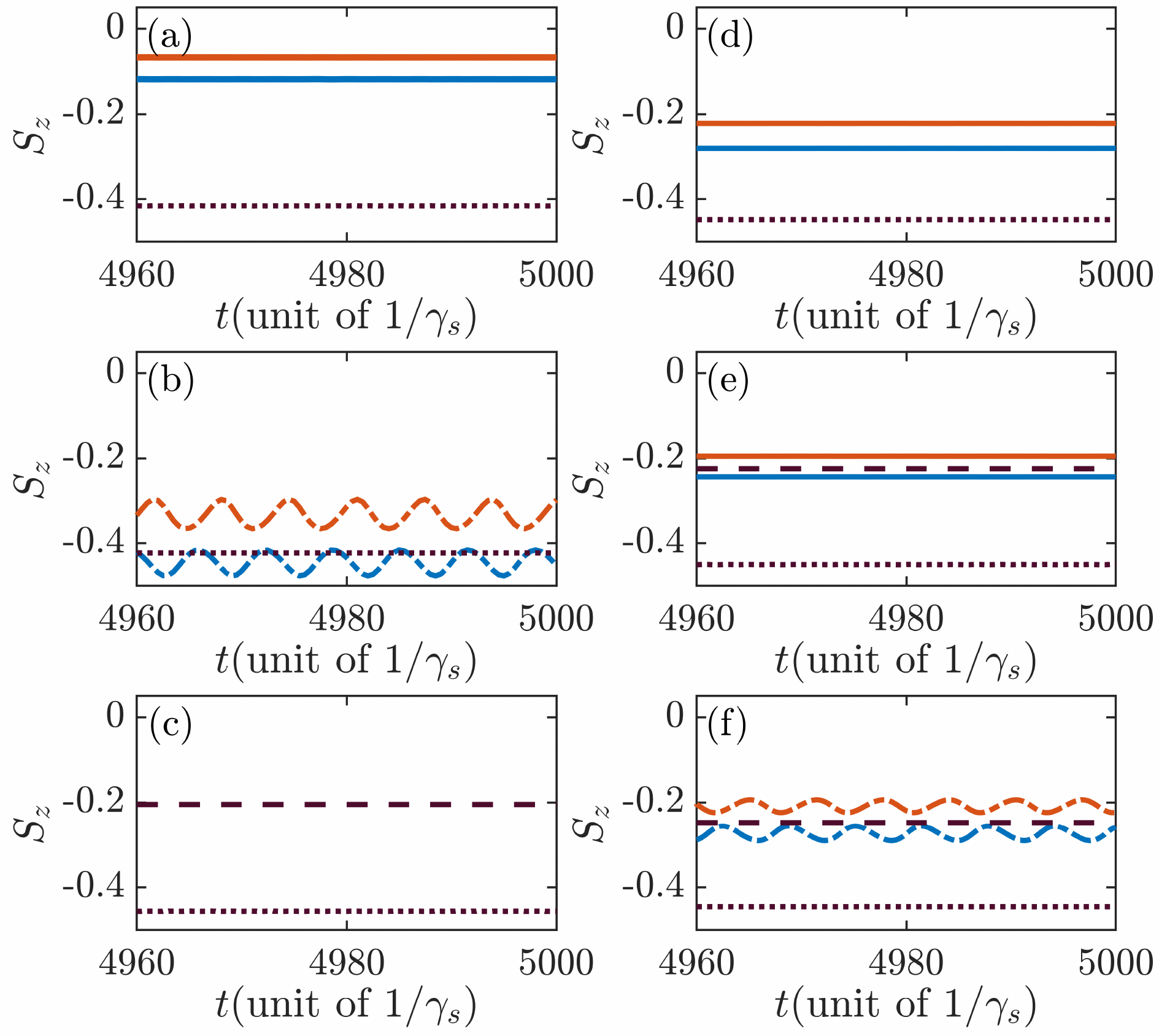}
	\caption{Dynamics for the bistable/multistable phases with the same parameters as in Fig. \ref{fig:N=2phase-V}(b). (a) AFM/ULE phase with $\Delta = -1.15$, $\Omega = 2$, (b) OSC/ULE phase with $\Delta = -4.14$, $\Omega = 0.56$, (c) ULE/UHE phase with $\Delta = -0.5$, $\Omega = 2.5$, (d) M1 phase with $\Delta = -2.6$, $\Omega = 1.01$, (e) AFM/OSC/ULE/UHE phase with $\Delta = -2.28$, $\Omega = 1.14$, (f) OSC/ULE/UHE phase with $\Delta = -2.51$, $\Omega = 1.07$. }
	\label{fig:dynamicsN=2phase}
\end{figure}

\section{More Examples of Population Dynamics in the MF regime}\label{Examples of Dynamics}
We simulate the dynamic evolution process to get some insight into the characteristics of different phases with nonlocal dissipation. Fig.~\ref{fig:dynamicsN=50interaction} shows the dynamics of the first six sites ($N = 50$) with different initial states in the long-time limit around $t = 5000$. Fig.~\ref{fig:dynamicsN=50interaction}(a) shows when $V_1 = 0$, $V_2 = 5$, the spins with different initial states evolve through time to reach the same steady state at the long-time limit. The atoms are almost in the lower state, which is in the ULE phase. Fig.~\ref{fig:dynamicsN=50interaction}(b) shows the negative interaction $V_2= -5$ drives the atoms from the lower state into the superposition state, which dynamics show ULE phase become UHE phase. Fig.~\ref{fig:dynamicsN=50interaction}(a), (c) and (d) have the same parameters but different interaction $V_1$. With the increase of the interaction $V_1$, the uniform phase gradually becomes nonuniform and enters the AFM phase. For the AFM phase, the system coexists in two stable steady states that evolve over time in which one has a higher population than the other. Fig.~\ref{fig:dynamicsN=50interaction}(d) shows the population in the OSC phase oscillates periodically in time as $V_1$ further increase.

For a single simulation, we typically obtain one phase. The bistable and multistable phases are found in different simulations. We consider different initial states to check for bistability. Fig.~\ref{fig:dynamicsN=2phase} shows examples of spin dynamics corresponding to the bistable and multistable phase regions in Fig.~\ref{fig:N=2phase-V}(b). The left panel represents the bistable phases and the right panel represents the multistable phases. In the bistable phase, both phases can coexist. The M2 phase show the existence of AFM and OSC phase (see Fig.~\ref{fig:dynamicsN=2phase}(e) and (f)).

\begin{figure}[htp!]
	\centering   \includegraphics[width=0.8\linewidth]{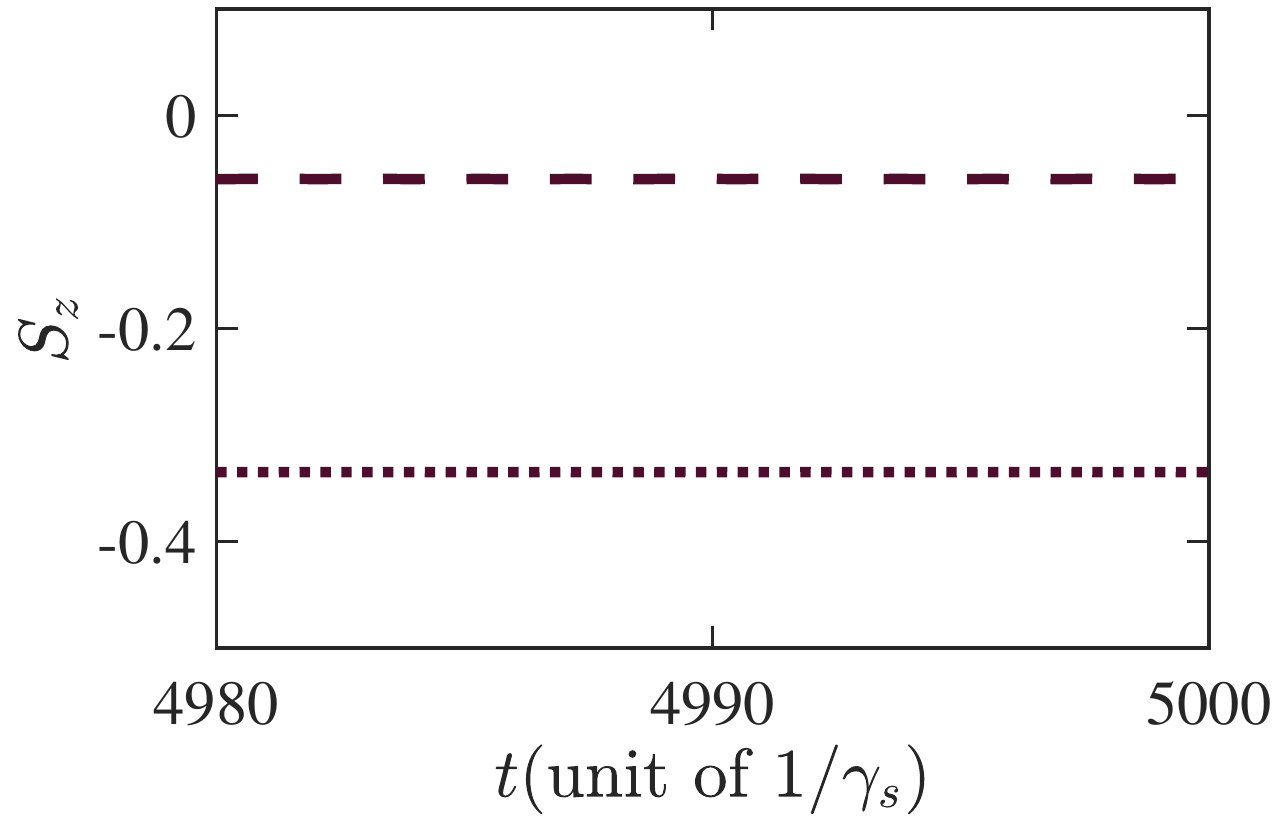}
	\caption{Dynamics simulations of the M2 phase. When $N=50$, only ULE/UHE phase exists. Different lines represent different simulations. Other Parameters: $V_1=V_2=5$, $\Delta=-2$, $\gamma_s=\gamma_m=0.5$.}
	\label{fig:dynamicsM2}
\end{figure}

As $N$ increases, only ULE/UHE phase stability exists. In the bistable ULE/UHE phase, on the other hand, all sites will have identical occupation, and hence all curves collapse to a single line. However, they could have either low occupation or high occupation, depending on the initial condition. In Fig.~\ref{fig:dynamicsM2}, we have shown two examples from different simulations where all sites have higher (lower) occupations, illustrating the bistable phase.

%

\end{document}